\documentclass[prd,aps,reprint,nofootinbib,superscriptaddress,floatfix,onecolumn,eqsecnum]{revtex4-2}
\usepackage{listings}
\usepackage{graphicx}
\usepackage{float}
\graphicspath{{figures/}}
\usepackage{dcolumn}
\usepackage{bm}
\usepackage{booktabs}
\usepackage{amsmath,amssymb}
\usepackage{framed}
\usepackage{xcolor}
\usepackage{physics}
\usepackage[colorlinks,linkcolor=blue]{hyperref}
\usepackage{orcidlink}
\usepackage{setspace}
\usepackage[normalem]{ulem}
\usepackage{empheq}
\usepackage{cancel}
\definecolor{orcidlogocol}{HTML}{A6CE39}
\newcommand{\orcid}[1]{\href{https://orcid.org/#1}{\textcolor[HTML]{A6CE39}{\aiOrcid}}}

\newcommand{\Mstar}{M_\ast}

\usepackage{etoolbox}

\makeatletter
\patchcmd{\frontmatter@abstract@produce}
  {\vskip200\p@\@plus1fil
   \penalty-200\relax
   \vskip-200\p@\@plus-1fil}
  {}
  {}
  {}
\makeatother
\allowdisplaybreaks
\makeatletter
\def\l@subsection#1#2{}
\makeatother

\begin{document}

\title{Global causality constraints in rotating scalar-tensor spacetimes}
\author{Bum-Hoon Lee \orcidlink{0009-0008-3322-2087}}
\email{bhl@sogang.ac.kr}
\affiliation{Department of Physics, Shanghai University, Shanghai, 200444, China}
\affiliation{Center for Quantum Spacetime, Sogang University, Seoul 121-742, Korea}
\affiliation{Department of Physics, Sogang University, Seoul 121-742, Korea}

\author{Nils A. Nilsson \orcidlink{0000-0001-6949-3956}}
\email{nilsson@ibs.re.kr}
\affiliation{Cosmology, Gravity and Astroparticle Physics Group, Center for Theoretical Physics of the Universe, Institute for Basic Science, Daejeon 34126, Korea}
\affiliation{LTE, Observatoire de Paris, Université PSL, CNRS, LNE, Sorbonne Universit\'e, 61 avenue de l'Observatoire, 75 014 Paris, France}

\author{Somyadip Thakur \orcidlink{0000-0003-2376-5906}}
\email{somyadip@sogang.ac.kr}
\affiliation{Center for Quantum Spacetime, Sogang University, Seoul 121-742, Korea}

\date{\today}

\begin{abstract}
Modified gravity is often formulated as an effective field theory (EFT), where higher-order corrections parametrize departures from General Relativity. We argue that such corrections should be constrained by the global causal structure of curved spacetime, in addition to the usual flat-space requirements such as positivity and unitarity. We propose that within the domain of validity of the EFT, the onset of closed timelike curves should not happen in a parametrically more accessible region than in the corresponding GR background. We test this diagnostic in the quadratic k-essence sector of scalar-tensor gravity. For stationary and axisymmetric spacetimes, the invariant test for closed axial orbits is the sign of the azimuthal component of the metric  \(g_{\varphi\varphi}\). We supplement this test by requiring a local time function in the space of Killing vectors. We apply these conditions to quadratic k-essence on Kerr--(A)dS backgrounds, with and without scalar charge. The zero-charge branch is exact Kerr--(A)dS, and we treat the charged branch perturbatively in scalar charge and in Hartle--Thorne slow rotation. Expanding for small spin \(\chi=a/(GM)\ll1\), frame dragging begins at \(\mathcal O(\chi)\), while the quadrupolar backreaction relevant for circular closed timelike curves enters at second order in both rotation and charge. We find that, in the truncation used here, any occurrence of \(g_{\varphi\varphi}<0\) also lies outside EFT control. A higher-order calculation or a fully nonlinear treatment is therefore needed. Finally, we discuss how quasinormal modes and black-hole echoes could probe such causal structure.
\end{abstract}

\maketitle

\section{Introduction}\label{sec:intro}

General relativity (GR) agrees with precision tests on many scales, from the Solar System to binary systems, and gravitational-wave detections now probe the nonlinear regime of compact-binary coalescence~\cite{Will:2014kxa,Bourgoin:2016ynf,Bourgoin:2017fpo,Fienga:2023ocw,LIGOScientific:2016aoc,LIGOScientific:2020ibl,KAGRA:2021vkt}.  Future ground- and space-based interferometers will sharpen these tests, both by increasing compact-object statistics and by improving sensitivity to small beyond-GR effects~\cite{Maggiore:2019uih,Reitze:2019iox,Barausse:2020rsu}.  At the same time, GR is not expected to be the final description of gravity: as a perturbative quantum field theory it is non-renormalisable, it admits classical curvature singularities, and in cosmology, parameter tensions and the cosmological constant problem remain active topics of research~\cite{Burgess:2003jk,Martin:2012bt,CosmoVerse:2025txj}. These problems motivate modifications of GR, whose consistency must be tested independently. 

A modified-gravity model must pass at least the same theoretical and observational tests as GR. One such test is causality. In flat-space EFT, analyticity, unitarity, and locality constrain Wilson coefficients and the allowed operators. Operators with the wrong sign can produce ghost modes, superluminal propagation, or time advances, obstructing the usual notion of a standard UV completion~\cite{Adams:2006sv,deRham:2017avq,CarrilloGonzalez:2022fwg}.  On curved backgrounds, causality is more subtle because the EFT cutoff, the local characteristic light cone, and the global causal structure of spacetime are tied together.  Recent work has therefore formulated infrared causality bounds using Shapiro time advances, time delays, and quasinormal-mode propagation on nontrivial geometries~\cite{deRham:2020zyh,Alberte:2020jsk,Tokuda:2020mlf,Melville:2024zjq}. 
In contrast to local light-cone statements, Closed Timelike Curves (CTCs) are a global causality problem. If a timelike curve closes on itself, chronological ordering is lost and the spacetime is not globally hyperbolic~\cite{Minguzzi:2019mbe}.

Energy conditions provide a related, but not equivalent, way to control gravitational causality.  Through the Raychaudhuri equation, the null convergence condition \(R_{\mu\nu}k^\mu k^\nu\geq0\), which follows from the null energy condition in GR, controls the focusing of hypersurface-orthogonal null congruences.  This input underlies many classical singularity and area theorems, and explains why violations of the null energy condition often appear in discussions of traversable wormholes, chronology violation, and other exotic geometries.  The converse is not true.  Local defocusing, or even superluminal characteristics, does not automatically produce a closed causal curve. Whether a CTC forms is a global question about the cone structure and the identifications of the spacetime.  Here we treat CTCs as a separate global consistency test for modified-gravity backgrounds.

Several exact GR solutions contain closed timelike curves, including rotating-cylinder spacetimes related to the Lanczos--van Stockum solution and the G\"odel universe~\cite{lanczos,vanStockum:1937zz,godel,tipler,Lobo:2008leb}. Many of these examples rely on idealised global structure or unphysical matter. Hawking's chronology protection conjecture expresses the expectation that the laws of physics prevent CTCs from forming in physically admissible situations~\cite{hawking,Visser:2002ua}.  We use this expectation as a criterion for modified gravity: a deviation from GR should not push CTCs into a region of parameter space where the EFT is meant to apply. We do not try to prove a general chronology-protection theorem, since CTCs can occur even in GR, though usually hidden behind horizons. Instead, we ask whether a given EFT deformation moves a chronology-violation threshold to a point that is easier to reach than in GR. We apply this idea to scalar-tensor gravity. Lovelock's theorem restricts purely metric, four-dimensional, second-order modifications of the Einstein equations, so scalar-tensor theories provide one of the minimal systematic routes beyond GR~\cite{Lovelock:1971yv,Lovelock:1972vz,Horndeski:1974wa,Kobayashi:2019hrl}.  We focus on the k-essence sector, where non-canonical scalar kinetic terms can modify both the background geometry and the characteristic cone of perturbations~\cite{Armendariz-Picon:2000ulo}.  Metric causality and field-characteristic causality need not coincide.  Superluminal characteristics in k-essence do not by themselves make a time machine.  We instead ask whether the metric, or the relevant effective metric, admits closed causal curves inside the controlled EFT domain~\cite{Liberati:2001sd,Babichev:2007dw}.  We test whether a k-essence deformation of a GR solution can move the onset of circular CTCs on a rotating background, and how any controlled onset would constrain the EFT coefficients.

We first illustrate the logic with a rotating cylinder and then apply it to quadratic k-essence on Kerr--(A)dS backgrounds. In the zero-charge branch the geometry is exact Kerr--(A)dS. In the charged branch we use the Hartle--Thorne hierarchy: the quadrupolar radial response is obtained in the leading even-parity slow-rotation sector and enters the circular diagnostic at second order in both rotation and scalar charge. We support the scalar profile by an external shift current and solve the reduced radial response equations. In the regular sourceless radial limit the quadrupolar response is absent, so the circular zeros of $g_{\varphi\varphi}$ studied below are not generated. With the external charge source included, we track when a chronology-violating region appears or disappears and relate this to the model parameters and to EFT control. Finally, we discuss gravitational-wave echoes as a possible observational signature, but only after imposing an additional reflective boundary on the chronology-violating surface.

Section~\ref{sec:tipler} sets up the diagnostic, Sec.~\ref{sec:kessence} defines the k-essence backgrounds, Sec.~\ref{sec:kessenceCTCs} gives the main results for different black-hole spins, Sec.~\ref{sec:echoes} discusses the conditional ringdown response, and Sec.~\ref{sec:disc} concludes.  We use units \(c=\hbar=1\) and metric signature \((-+++)\).

\section{Closed timelike curves around a rotating cylinder}
\label{sec:tipler}

We begin with a locally flat geometry that shows how rotation can generate closed timelike curves. Minkowski space in cylindrical coordinates is
\begin{equation}
    ds^2=-dt^2+dr^2+r^2d\varphi^2+dz^2 .
\end{equation}
Introducing a constant off-diagonal \(dt\,d\varphi\) term gives the rotating-cylinder line element
\begin{equation}
    ds^2
    =
    -dt^2
    +2\omega\,dt\,d\varphi
    +(r^2-\omega^2)d\varphi^2
    +dr^2+dz^2 ,
    \label{eq:rotcyl}
\end{equation}
which is locally a vacuum solution. This differs from the van Stockum and Tipler rotating-cylinder solutions, which are sourced by dust~\cite{vanStockum:1937zz,tipler}. We use Eq.~\eqref{eq:rotcyl} to isolate the causal role of the azimuthal metric component without adding matter at the background level. The coordinate \(T\) introduced below is not globally single-valued, and the surface \(r=0\) is not an ordinary regular polar axis of the quotient once the axial orbit becomes timelike.

Closed azimuthal curves are generated by the Killing vector \(\partial_\varphi\). Since their squared norm is
    $g_{\varphi\varphi}=r^2-\omega^2$
they become timelike whenever
$g_{\varphi\varphi}<0$ and therefore when $r<|\omega|$.
The surface
\begin{equation}
    r_{\rm ctc}^{(0)}=|\omega|
\end{equation}
is therefore the chronology horizon for circular azimuthal loops in the cylinder. The line element in Eq.~\eqref{eq:rotcyl} is locally flat: the change of variables
\begin{equation}
    T=t-\omega\varphi
\end{equation}
brings it to
\begin{equation}
    ds^2=-dT^2+dr^2+r^2d\varphi^2+dz^2 .
\end{equation}
The transformation is not globally single-valued when \(\varphi\sim\varphi+2\pi\), because \(T\to T-2\pi\omega\) after one circuit. The CTC region is therefore a global effect of the helical identification, shown in the original geometry by the sign of \(g_{\varphi\varphi}=r^2-\omega^2\). In the following formulas we take \(\omega>0\) without loss of generality; restoring both orientations amounts to replacing \(\omega\) by \(|\omega|\) in the unperturbed root.

We use this background as a warmup for the backreaction analysis used later in the paper. We add a minimally coupled massless scalar field,
\begin{equation}
    S
    =
    \int d^4x\,\sqrt{-g}
    \left[
        \frac{R}{2\kappa}
        -\frac12(\partial_\mu\psi)^2
    \right],
    \label{eq:secIII_action}
\end{equation}
and treat the scalar amplitude perturbatively. We therefore write
\begin{equation}
    \psi(r)=q\,\tilde{\psi}(r)+\mathcal O(q^3),
    \qquad
    g_{\mu\nu}
    =
    g^{(0)}_{\mu\nu}
    +
    \varepsilon h_{\mu\nu}
    +
    \mathcal O(\varepsilon^2),
    \qquad
    \varepsilon\equiv \kappa q^2 .
    \label{eq:cylinder_scalar_ansatz}
\end{equation}
The detailed solution of the linearised Einstein equations is given in Appendix~\ref{app:cyl_details}.  For the profile
\begin{equation}
    \tilde\psi(r)=\frac12\ln\left(\frac{r^2+r_0^2}{r_0^2}\right),
    \qquad
    \Box\psi=\frac{2q r_0^2}{(r^2+r_0^2)^2},
    \label{eq:scalar_profile}
\end{equation}
the scalar is externally supported near the core and is not a sourceless solution of Eq.~\eqref{eq:secIII_action}.  In the coordinate \(t\), the first-order azimuthal component has the form
\begin{equation}
    g_{\varphi\varphi}(r)
    =r^2-\omega^2+
    \varepsilon\left[c(r)-2\omega B_0-\omega^2a(r)\right]
    +\mathcal O(\varepsilon^2),
    \label{eq:gphiphi_full}
\end{equation}
where \(a\), \(c\), and the constant \(B_0\) are defined in Appendix~\ref{app:cyl_details}.  Expanding the root of \(g_{\varphi\varphi}=0\) around \(r=\omega\) gives
\begin{equation}
    r_{\rm ctc}
    =
    \omega-
    \varepsilon\frac{c(\omega)-2\omega B_0-\omega^2a(\omega)}{2\omega}
    +\mathcal O(\varepsilon^2),
    \label{eq:rctc_B0}
\end{equation}
for \(\omega>0\), with \(\omega\to|\omega|\) for arbitrary orientation.  A positive value of the root shift \(r_{\rm ctc}-\omega\) enlarges the circular-CTC region, while a negative value delays its onset.

\begin{figure}[t]
  \centering \includegraphics[width=0.7\textwidth]{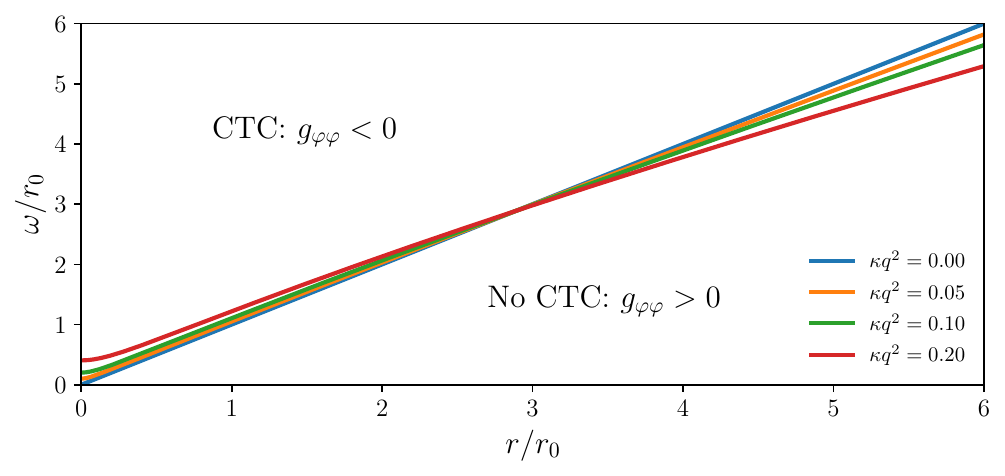}
\caption{
Circular-CTC regions for the locally flat rotating-cylinder quotient using the coordinate-basis sign of \(g_{\varphi\varphi}\) at fixed quotient time \(t\). The boundary \(g_{\varphi\varphi}=0\) marks the circular chronology horizon. The plotted scalar correction uses the homogeneous-mode choice described in Appendix~\ref{app:cyl_details} with the growing \(B_2\) mode removed and, for the displayed boundary, the constant frame mode set to \(B_0=0\); a different matching choice shifts the boundary according to Eq.~\eqref{eq:rctc_B0}.
}
\label{fig:cylinderCTC}
\end{figure}

Figure~\ref{fig:cylinderCTC} shows this shift in the \((r/r_0,\omega/r_0)\) plane. The unperturbed chronology horizon is the straight line \(r=\omega\). Once the scalar is included, the \(\mathcal O(\varepsilon)\) correction in Eq.~\eqref{eq:rctc_B0} deforms the boundary. The same mechanism  applies in  the k-essence model: backreaction changes the metric, the metric changes the location of the chronology horizon, and, once the matter sector is specified, the requirement that CTCs not move into the physical exterior region becomes a constraint on the underlying theory.

For k-essence black holes, we supplement the circular-CTC test with a determinant diagnostic for the Killing sector, which tracks the loss of a local time function in the stationary two-Killing plane. We do not identify every timelike helical Killing direction with a CTC. Such a direction becomes a closed timelike curve only after an additional global identification closes the orbit.

\subsection{Causality in stationary and axisymmetric geometry}
\label{sec:CTCcriterion}
In the zero-charge calculation below we measure rotation by
\begin{equation}
    \chi\equiv \frac{a}{GM},
    \label{eq:spin_chi_def}
\end{equation}
where $a$ is the Kerr spin parameter.  In the zero-charge branch the Kerr--(A)dS expressions are exact in \(\chi\).  In the charged branch, the same dimensionless spin \(\chi\) is used as the Hartle--Thorne slow-rotation parameter, so we assume \(\chi\ll1\).  The frame-dragging correction is odd in spin and starts at \(\mathcal O(\chi)\), while the even-parity quadrupolar correction relevant for \(g_{\varphi\varphi}\) starts at \(\mathcal O(\chi^2)\).  We therefore keep only the leading \(\mathcal O(\chi^2\varepsilon_q^2)\) sector in the charged circular diagnostic and omit higher-spin terms. We restrict to the circular, stationary-axisymmetric metric class used throughout the paper, so the metric has no \((t,\varphi)\)-\((r,\theta)\) cross terms. This assumption holds for the exact Kerr--(A)dS metric and for the Hartle--Thorne ansatz with the source included below; without it, the Killing condition \(\Delta_{\rm K}<0\) is not equivalent to \(g^{tt}<0\) for the full inverse metric. Let such a spacetime admit two commuting Killing fields
\begin{equation}
    \xi_{(t)}^\mu=(\partial_t)^\mu,
    \qquad
    \xi_{(\varphi)}^\mu=(\partial_\varphi)^\mu,
\end{equation}
with \(\varphi\sim\varphi+2\pi\). Circular azimuthal CTCs occur when the closed orbits of \(\xi_{(\varphi)}\) become timelike, i.e. when
\begin{equation}
    g(\xi_{(\varphi)},\xi_{(\varphi)})
    \equiv g_{\varphi\varphi}<0 .
    \label{eq:circular_ctc_condition}
\end{equation}
The scalar \(g(\xi_{(\varphi)},\xi_{(\varphi)})\) is invariant once the axial Killing generator and its periodic identification are fixed. Coordinate transformations can move the coordinate position of the root, but they cannot change whether the axial Killing orbit itself is timelike. The metric induced on the two-dimensional space spanned by the Killing fields is
\begin{equation}
    g_{AB}
    \equiv
    \begin{pmatrix}
        g_{tt} & g_{t\varphi} \\
        g_{t\varphi} & g_{\varphi\varphi}
    \end{pmatrix},
    \qquad A,B\in\{t,\varphi\},
    \label{eq:killing_block}
\end{equation}
and we define
\begin{equation}
    \Delta_{\rm K}
    \equiv
    \det(g_{AB})
    =
    g_{tt}g_{\varphi\varphi}-g_{t\varphi}^{2} .
    \label{eq:DeltaK_def}
\end{equation}
When \(g_{\varphi\varphi}>0\), the condition \(\Delta_{\rm K}<0\) states that the Killing space has Lorentzian signature; in the circular metric class just specified this is equivalently \(g^{tt}<0\) for the full metric. In this case \(t\) is a local time function for the Killing sector. In particular, for
\begin{equation}
    K(\Omega)=\partial_t+\Omega\partial_\varphi,
\end{equation}
one has
\begin{equation}
    g(K,K)
    =
    g_{tt}+2\Omega g_{t\varphi}+\Omega^2 g_{\varphi\varphi},
\end{equation}
with minimum
\begin{equation}
    g(K,K)_{\rm min}
    =
    \frac{\Delta_{\rm K}}{g_{\varphi\varphi}}
    \qquad
    (g_{\varphi\varphi}>0).
    \label{eq:Killing_norm_min}
\end{equation}
Thus \(\Delta_{\rm K}<0\) guarantees that some Killing direction is timelike, as in ordinary stationary Minkowski or Kerr exteriors. The integral curves of \(K(\Omega)\) are not closed in a standard stationary spacetime because one circuit in \(\varphi\) changes \(t\) by \(2\pi/\Omega\). They become helical CTCs only if the global quotient also identifies time, or if some other global condition closes the orbit. We therefore use the causality criteria
\begin{equation}
    g_{\varphi\varphi}>0,
    \qquad
    \Delta_{\rm K}<0,
    \label{eq:InvariantCTCcriterion}
\end{equation}
in the exterior domain. The first inequality excludes circular CTCs generated by the axial Killing orbit. The second rules out the loss of a local time function in the Killing sector and separates the chronology diagnostic from horizons or coordinate-boundary pathologies.
Residual transformations that preserve \(\partial_t\), \(\partial_\varphi\), and the \(2\pi\)-periodic axial identification shift the coordinate location of the roots of \(g_{\varphi\varphi}\) and \(\Delta_{\rm K}\). Transformations that mix \(t\) and \(\varphi\) change the global interpretation and are not part of this residual gauge freedom. In the zero-charge Kerr--(A)dS case, \(\Delta_{\rm K}=0\) reduces analytically to a horizon or angular-boundary condition, so \(g_{\varphi\varphi}=0\) gives the circular-CTC test after the usual domain restrictions are imposed.

\section{Quadratic k-essence backgrounds}
\label{sec:kessence}

In the zero-charge k-essence case, we treat the Kerr--(A)dS background geometry and the chronology criteria exactly in the spin parameter. In the charged case, we use a double perturbative expansion in scalar charge and Hartle--Thorne slow rotation.  We keep the leading even-parity quadrupolar response, \(\mathcal O(\chi^2\varepsilon_q^2)\), and leave higher-spin quadrupoles \(\mathcal O(\chi^4\varepsilon_q^2)\) and higher-charge corrections \(\mathcal O(\varepsilon_q^4)\) for future work.  For the kinetic profile used below, the radial functions are fixed by the reduced Hartle--Thorne response equations. We supplement the source-free action by an external shift-current source only in the small-charge branch,
\begin{align}
    S
    &=
    \int d^4x\,\sqrt{-g}
    \left[
        \frac{R}{2\kappa}
        +
        P(X)
    \right]
    +
    S_{\rm src},
    \label{eq:kessenceL}
    \\
    S_{\rm src}
    &=
    \int d^4x\,\sqrt{-g}\,
    J_{\rm ext}^{\mu}\nabla_\mu\phi
    +
    S_{\rm ext}[g,J_{\rm ext},\Psi_{\rm ext}],
    \label{eq:source_action}
\end{align}
where
\begin{equation}
    X
    \equiv
    -\frac12 g^{\mu\nu}
    \partial_\mu\phi\,\partial_\nu\phi .
\end{equation}
The vector $J_{\rm ext}^{\mu}$ represents the effective external sector that supports the prescribed scalar current.  The derivative coupling in Eq.~\eqref{eq:source_action} preserves the shift symmetry of the k-essence scalar.  We recover the zero-charge branch and the sourceless radial limit by setting $J_{\rm ext}^{\mu}=0$.  Varying with respect to $\phi$ gives
\begin{equation}
    \nabla_\mu
    \left(
        P_X\nabla^\mu\phi
    \right)
    =
    \nabla_\mu J_{\rm ext}^{\mu}
    \equiv
    \mathcal J_{\rm ext},
    \label{eq:kessence_scalar_sourced}
\end{equation}
while the metric equation is
\begin{equation}
    G_{\mu\nu}
    =
    \kappa
    \left(
        T^{(\phi)}_{\mu\nu}
        +
        T^{\rm ext}_{\mu\nu}
    \right),
    \qquad
    T^{(\phi)}_{\mu\nu}
    =
    P_X\partial_\mu\phi\partial_\nu\phi
    +
    g_{\mu\nu}P(X),
    \label{eq:kessenceeom}
\end{equation}
with $P_X\equiv\partial P/\partial X$.  The current fixes the scalar equation but not the stress tensor of the external sector.  Appendix~\ref{app:exact_source_closure} gives the residual stress needed to close the Einstein equations in the nonzero-charge case.

We consider the quadratic model
\begin{equation}
    P(X)=\alpha X+\beta X^2,
    \label{eq:P(X)def}
\end{equation}
where $\alpha$ is dimensionless and $[\beta]=M^{-4}$ in units $c=\hbar=1$.  With the canonical four-dimensional scalar mass dimension $[\phi]=M$, one has $[X]=M^4$ and $[q]=M^2$.  We write
\begin{equation}
    \beta=\frac{\tilde\beta}{\Mstar^4},
    \qquad
    B\equiv\frac{\tilde\beta}{\alpha^2}
    =\Mstar^4\frac{\beta}{\alpha^2},
    \qquad
    s_\alpha\equiv\operatorname{sgn}(\alpha),
    \label{eq:simplified_B_def}
\end{equation}
where $\Mstar$ is the EFT scale.  We therefore use $B$ as the dimensionless coupling in the main text. After rescaling the scalar by $\sqrt{|\alpha|}$, the normalized kinetic variable is $X_c\equiv |\alpha|X$, and the quadratic coefficient is $B/\Mstar^4$.  We denote the source amplitude used in the plots by
\begin{equation}
    C_X\equiv |\alpha|\tilde C_X.
    \label{eq:CX_simple_def}
\end{equation}
The constant-$X$ case of the quadratic theory is defined by
\begin{equation}
    P_X
    =
    \alpha+2\beta X
    =
    0,
\end{equation}
so that
\begin{equation}
    X_\star
    =
    -\frac{\alpha}{2\beta}.
    \label{eq:Xstar_solution}
\end{equation}
In the normalized variable $X_c=|\alpha|X$, this becomes
\begin{equation}
    X_{c\star}
    =
    -\frac{s_\alpha\Mstar^4}{2B},
    \qquad
    P(X_\star)
    =
    -\frac{\Mstar^4}{4B},
    \qquad
    \Lambda_{\rm eff}
    =
    \frac{\kappa\Mstar^4}{4B} .
    \label{eq:Xstar_B_variables}
\end{equation}
The static spacelike constant-$X$ branch used in this paper therefore requires
\begin{equation}
    -2X_\star=\frac{\alpha}{\beta}>0,
    \qquad
    -2X_{c\star}=\frac{s_\alpha\Mstar^4}{B}>0,
    \label{eq:static_branch_reality_main}
\end{equation}
so $s_\alpha B>0$.  Thus $s_\alpha=+1$, $B>0$ is dS-like, while $s_\alpha=-1$, $B<0$ is AdS-like.  The negative-$\alpha$ branch contains ghosts around the trivial $X=0$ vacuum because $P_X(0)=\alpha<0$. Moreover, the constant-$X$ branch itself sits at $P_X=0$, where the k-essence scalar principal symbol is degenerate unless additional operators or external degrees of freedom restore hyperbolicity. We therefore only include the negative-$\alpha$ sector for completeness.

The EFT expansion is controlled by requiring the black-hole length scale to be larger than the cutoff scale,
\begin{equation}
    GM
    \gg
    \Mstar^{-1}.
\end{equation}
For the small-charge branch we also take
\begin{equation}
    \varepsilon_q
    \equiv
    q(GM)^2
    \ll 1,
    \qquad
    q
    \ll
    \Mstar^2 .
    \label{eq:epsilon_kessence_def}
\end{equation}
The first condition controls the backreaction, while the second keeps the charge within the EFT regime. The constant-$X$ background itself must also lie below the cutoff if the branch is to be interpreted as a Wilsonian configuration, so we take
\begin{equation}
    \frac{|X_{c\star}|}{\Mstar^4}
    =
    \frac{1}{2|B|}
    \ll 1 .
    \label{eq:Xstar_cutoff_condition}
\end{equation}
Parameter points that violate Eq.~\eqref{eq:Xstar_cutoff_condition} can still be used to diagnose the $P(X)$ function, but not as controlled low-gradient EFT points.

\subsection{Exact zero-charge Kerr--(A)dS branch}
\label{sec:kessencezerocharge}

In the zero-charge branch, $P_X(X_\star)=0$.  The stress tensor reduces to
\begin{equation}
    T_{\mu\nu}^{(0)}
    =
    P(X_\star)g_{\mu\nu}^{(0)},
    \qquad
    P(X_\star)
    =
    -\frac{\alpha^2}{4\beta}
    =
    -\frac{\Mstar^4}{4B}.
\end{equation}
The Einstein equation becomes
\begin{equation}
    G_{\mu\nu}[g^{(0)}]
    +
    \Lambda_{\rm eff}g_{\mu\nu}^{(0)}
    =
    0,
    \label{eq:kessence_zerocharge_Einsteineqs}
\end{equation}
with \(\Lambda_{\rm eff}\) defined in Eq.~\eqref{eq:Xstar_B_variables}.  Thus the zero-charge branch is exactly Kerr--(A)dS.

In Boyer--Lindquist coordinates we use the metric
\begin{equation}
    ds^2
    =
    -\frac{\Delta_r}{\rho^2}
    \left(
        dt-\frac{a\sin^2\theta}{\Xi}d\varphi
    \right)^2
    +
    \frac{\rho^2}{\Delta_r}dr^2
    +
    \frac{\rho^2}{\Delta_\theta}d\theta^2
    +
    \frac{\Delta_\theta\sin^2\theta}{\rho^2}
    \left(
        adt-\frac{r^2+a^2}{\Xi}d\varphi
    \right)^2,
    \label{eq:Kerr-AdS_metric}
\end{equation}
where
\begin{equation}
\begin{aligned}
    \rho^2
    &=
    r^2+a^2\cos^2\theta,
    &
    \Xi
    &=
    1+\frac{\Lambda_{\rm eff}}{3}a^2,
    \\
    \Delta_r
    &=
    (r^2+a^2)
    \left(
        1-\frac{\Lambda_{\rm eff}}{3}r^2
    \right)
    -2GMr,
    &
    \Delta_\theta
    &=
    1+\frac{\Lambda_{\rm eff}}{3}a^2\cos^2\theta .
\end{aligned}
\label{eq:kerrads_functions}
\end{equation}
This form is exact in the spin parameter $a$.  The coordinate domain is part of the background definition used in the chronology scan.  We impose \(\Xi>0\) and \(\Delta_\theta>0\), so that the angular sector and the AdS conformal boundary are well defined.  For $\Lambda_{\rm eff}>0$, we restrict to the rotating static patch between the outer black-hole horizon and the cosmological horizon.  For $\Lambda_{\rm eff}\le0$, we use the exterior of the outer black-hole horizon, again subject to \(\Xi>0\) and \(\Delta_\theta>0\).

We introduce the remaining dimensionless variables, using the spin parameter \(\chi\) already defined in Eq.~\eqref{eq:spin_chi_def},
\begin{equation}
    \tau\equiv\frac{t}{GM},
    \qquad
    x\equiv\frac{r}{GM},
    \qquad
    \lambda\equiv\Lambda_{\rm eff}(GM)^2.
    \label{eq:x_chi_lambda_def}
\end{equation}
In these variables,
\begin{equation}
\begin{aligned}
    \bar\rho^2
    &=
    x^2+\chi^2\cos^2\theta,
    &
    \Delta_x
    &=
    (x^2+\chi^2)
    \left(
        1-\frac{\lambda x^2}{3}
    \right)
    -2x,
    \\
    \Delta_\theta
    &=
    1+\frac{\lambda\chi^2}{3}\cos^2\theta,
    &
    \Xi
    &=
    1+\frac{\lambda\chi^2}{3}.
\end{aligned}
\label{eq:dimless_kerrads_functions}
\end{equation}
The dimensionless metric $d\bar s^2\equiv ds^2/(GM)^2$ is
\begin{equation}
    d\bar s^2
    =
    -\frac{\Delta_x}{\bar\rho^2}
    \left(
        d\tau-\frac{\chi\sin^2\theta}{\Xi}d\varphi
    \right)^2
    +
    \frac{\bar\rho^2}{\Delta_x}dx^2
    +
    \frac{\bar\rho^2}{\Delta_\theta}d\theta^2
    +
    \frac{\Delta_\theta\sin^2\theta}{\bar\rho^2}
    \left(
        \chi d\tau-\frac{x^2+\chi^2}{\Xi}d\varphi
    \right)^2 .
    \label{eq:dimless_kerrads_metric}
\end{equation}
The exact Killing components are
\begin{align}
    \bar g_{\tau\tau}
    &=
    \frac{-\Delta_x+\chi^2\Delta_\theta\sin^2\theta}{\bar\rho^2},
    \label{eq:exact_gtautau}
    \\
    \bar g_{\tau\varphi}
    &=
    \frac{\chi\sin^2\theta}{\bar\rho^2\Xi}
    \left[
        \Delta_x-\Delta_\theta(x^2+\chi^2)
    \right],
    \label{eq:exact_gtauphi}
    \\
    \bar g_{\varphi\varphi}
    &=
    \frac{\sin^2\theta}{\bar\rho^2\Xi^2}
    \left[
        \Delta_\theta(x^2+\chi^2)^2
        -
        \chi^2\sin^2\theta\,\Delta_x
    \right].
    \label{eq:exact_gphiphi_dimless}
\end{align}
These expressions are dimensionless; the physical components involving $t$ differ by the corresponding powers of $GM$, but their signs and the sign of the determinant are unchanged.  The determinant identity is exact,
\begin{equation}
    \bar\Delta_{\rm K}^{\rm Kerr-(A)dS}
    \equiv
    \bar g_{\tau\tau}\bar g_{\varphi\varphi}
    -
    \bar g_{\tau\varphi}^2
    =
    -\frac{\Delta_x\Delta_\theta\sin^2\theta}{\Xi^2}.
    \label{eq:kerrads_det_identity}
\end{equation}
Thus $\Delta_{\rm K}=0$ identifies a horizon, the rotation axis, or an angular-domain breakdown in the zero-charge branch; it is not an independent helical-CTC threshold.

The exact azimuthal component can be written in the positive form
\begin{align}
    \Delta_\theta(x^2+\chi^2)^2
    -
    \chi^2\sin^2\theta\,\Delta_x
    &=
    (x^2+\chi^2)\bar\rho^2\Xi
    +
    2x\chi^2\sin^2\theta .
    \label{eq:kerrads_gphiphi_positive_identity_dimless}
\end{align}
For $x>0$, $M>0$, $0<\theta<\pi$, and $\Xi>0$, the right-hand side is positive.  Therefore
\begin{equation}
    \bar g_{\varphi\varphi}>0
    \label{eq:zerocharge_no_circular_ctc}
\end{equation}
everywhere in the usual exterior or dS static-patch domain.  The zero-charge branch can show circular CTCs only if one extends the Kerr--AdS geometry outside that domain, for example by allowing $\Xi\le0$.  These cases are the familiar rotating-AdS boundary pathology, which we display below.

\subsubsection{\texorpdfstring{Nonstandard AdS extension with $\Xi\le0$}{Nonstandard AdS extension with Xi <= 0}}
\label{sec:nonstandard_ads_zero_charge}

The positivity statement above used $\Xi>0$.  If the Kerr--AdS metric is continued past that domain condition, the azimuthal numerator
\begin{equation}
    \mathcal N_{\varphi}(x,\theta)
    =
    \Delta_\theta(x^2+\chi^2)^2
    -
    \chi^2\sin^2\theta\,\Delta_x
    \label{eq:Nphi_ctc_def_main}
\end{equation}
controls the sign of $g_{\varphi\varphi}$ away from the axis and away from the singular normalization $\Xi=0$.  On the equatorial plane,
\begin{equation}
    \mathcal N_{\varphi,{\rm eq}}(x)
    =
    (x^2+\chi^2)^2
    -
    \chi^2\Delta_x
    =
    x\left[\Xi x(x^2+\chi^2)+2\chi^2\right].
    \label{eq:Nphi_equatorial_path_ads}
\end{equation}
A nontrivial positive equatorial root must satisfy
\begin{equation}
    \Xi x_{\rm CTC}\left(x_{\rm CTC}^2+\chi^2\right)+2\chi^2=0,
    \label{eq:path_ads_equatorial_root}
\end{equation}
and for $\chi\ne0$ and $x_{\rm CTC}>0$, Eq.~\eqref{eq:path_ads_equatorial_root} requires
$\Xi<0$, leading to $\lambda< -3/\chi^2$ and
\begin{equation}
    \frac{\alpha^2}{\beta}<-\frac{12}{\kappa(GM)^2\chi^2} .
    \label{eq:path_ads_lambda_micro_bound}
\end{equation}
Equivalently, with
\begin{equation}
    C_\Lambda\equiv \kappa\Mstar^4(GM)^2,
\end{equation}
one has
$-C_\Lambda\chi^2/12<B<0$.
The large-$x$ behavior is
\begin{equation}
    \mathcal N_{\varphi,{\rm eq}}(x)
    \sim
    \Xi x^4,
    \qquad x\rightarrow\infty .
    \label{eq:path_ads_asymptotic_sign}
\end{equation}
Thus, when $\Xi<0$, $g_{\varphi\varphi}<0$ at sufficiently large radius, while near $x=0^+$ one has $\mathcal N_{\varphi,{\rm eq}}=2\chi^2 x+\mathcal O(x^2)>0$.  The positive root is therefore the inner edge of an outward-extending circular-CTC region, not a region confined near the horizon. The limiting case $\Xi=0$ is even less like an ordinary branch: the Boyer--Lindquist angular normalization and the AdS boundary frame become degenerate, the nontrivial equatorial root is pushed to infinity, and the metric no longer defines the usual Kerr--AdS boundary-value problem.  We classify the $\Xi\le0$ continuation as a Kerr--AdS domain pathology rather than a new k-essence microphysical bound. In the figures we label it as a nonstandard AdS extension and do not count it as a controlled circular-CTC region from the sourced branch.

\subsection{Hartle--Thorne sourced small-charge case}
\label{sec:nzflux}
For non-zero charge, we use the Hartle--Thorne expansion.  The circular component receives its leading quadrupolar correction at \(\mathcal O(\chi^2\varepsilon_q^2)\).  We obtain the radial functions from the reduced slow-rotation response system in Appendix~\ref{app:nzflux_technical}, using a kinetic profile supported by an external current.  This construction solves the Einstein equations within the reduced Hartle--Thorne sector.

We treat the charged diagnostic sector perturbatively in the charge parameter and write
\begin{equation}
    \phi(t,r,\theta)
    =
    \phi_\star(t,r)
    +
    \varepsilon_q^2\Phi(r,\theta)
    +
    \mathcal O(\varepsilon_q^4),
    \qquad
    \phi_\star(t,r)=\phi_0(r)+qt,
    \label{eqn:nzcharge_exact_spin}
\end{equation}
where $\varepsilon_q$ is defined in Eq.~\eqref{eq:epsilon_kessence_def}.  Since the stress tensor is quadratic in scalar derivatives, the leading charge backreaction enters at $\mathcal O(\varepsilon_q^2)$.
The full nonzero-charge Hartle--Thorne expansion can contain monopole terms at \(\mathcal O(\varepsilon_q^2)\), a charge correction to frame dragging at \(\mathcal O(\chi\varepsilon_q^2)\), and even-parity quadrupolar terms at \(\mathcal O(\chi^2\varepsilon_q^2)\).  The sign of \(g_{\varphi\varphi}\) is affected first by the even-parity quadrupolar sector, so we write
\begin{equation}
    \left.\frac{ds^2}{(GM)^2}\right|_{\rm circ}
    =
    d\bar s^2_{\rm Kerr-(A)dS}(\chi,\lambda)
    +
    \varepsilon_q^2\chi^2 h^{(q,2)}_{\mu\nu}(x,\theta;\lambda)dx^\mu dx^\nu
    +
    \mathcal O(\varepsilon_q^2\chi^4,\varepsilon_q^4\chi^2),
    \label{eq:ht_charge_metric_expansion}
\end{equation}
for the circular component and its companion even-parity metric functions.  Here $d\bar s^2_{\rm Kerr-(A)dS}$ is Eq.~\eqref{eq:dimless_kerrads_metric}. The omitted monopole and frame-dragging terms can affect other observables and, in particular, the determinant diagnostic $\Delta_K$. Those corrections lie beyond the approximation used here. At fully nonlinear order one would determine the full two-dimensional function $h^{(q)}_{\mu\nu}(x,\theta)$.  In the calculation below the external current fixes the scalar profile, and we obtain the retained radial response from the reduced Hartle--Thorne equations summarized in Appendix~\ref{app:nzflux_technical}.

The kinetic term is expanded as
\begin{equation}
    X(x,\theta)
    =
    X_\star
    +
    \varepsilon_q^2\Mstar^4
    \left[
        \mathcal X_{2,0}(x)
        +
        \mathcal X_{2,2}(x)P_2(\cos\theta)
    \right]
    +
    \mathcal O(\varepsilon_q^4\Mstar^4).
    \label{eq:Xexpand_nz}
\end{equation}
Since $P_X(X_\star)=0$, we have
\begin{equation}
    P_X(X)
    =
    2\tilde\beta\varepsilon_q^2
    \left[
        \mathcal X_{2,0}(x)
        +
        \mathcal X_{2,2}(x)P_2(\cos\theta)
    \right]
    +
    \mathcal O(\varepsilon_q^4),
    \label{eq:PXexpand_nz}
\end{equation}
but $P$ itself remains
$
    P(X)
    =
    P(X_\star)
    +
    \mathcal O(\varepsilon_q^4),
$
because the linear correction in $X-X_\star$ is multiplied by $P_X(X_\star)$.  Therefore, at $\mathcal O(\varepsilon_q^2)$, the scalar-sector stress tensor separates into the zero-charge vacuum-energy piece and a charge-induced source as
\begin{equation}
    T^{(\phi)}_{\mu\nu}
    =
    P(X_\star)g_{\mu\nu}
    +
    2\tilde\beta\varepsilon_q^2
    \left[
        \mathcal X_{2,0}(x)
        +
        \mathcal X_{2,2}(x)P_2(\cos\theta)
    \right]
    \nabla_\mu\phi_\star\nabla_\nu\phi_\star
    +
    \mathcal O(\varepsilon_q^4).
    \label{eq:Tsplit_nz_correct}
\end{equation}
With the external source included, the quadrupolar contributions are nonzero
\begin{equation}
    \mathcal X_{2,2}\neq0,
    \qquad
    \Pi_2\neq0,
    \qquad
    S_K\neq0.
    \label{eq:sourced_nonzero_conditions_main}
\end{equation}
The source is not inserted by hand only at the level of the scalar equation; it follows from the action term in Eq.~\eqref{eq:source_action}.  In the notation of Eq.~\eqref{eq:kessence_scalar_sourced}, we choose the external current so that
$
    \nabla_\mu J_{\rm ext}^{\mu}
    =
    \mathcal J_{\rm ext}
$
and the prescribed nonzero profile $\mathcal X_{2,2}(x)$ satisfies the sourced scalar equation.  Appendix~\ref{app:nzflux_technical} gives the explicit current, the AdS exterior boundary problem, and the dS static-patch boundary problem.
By contrast, we do not include the corresponding sourceless radial branch. If $J_{\rm ext}^{\mu}=0$, the quadrupolar scalar equation has a first integral.  Regularity at a nonextremal black-hole horizon sets the integration constant to zero, giving
$
    \mathcal X_{2,2}=0,
    \,
    \Pi_2=0,
    \,
    S_K=0.
$
With homogeneous quadrupolar boundary conditions this removes the charge-induced response.  The regular sourceless radial limit therefore does not affect the circular-CTC diagnostic at $\mathcal O(\varepsilon_q^2)$, and we drop it from the analysis.

For the minimal leading-spin diagnostic, we write the relevant \(\mathcal O(\chi^2\varepsilon_q^2)\) components of the charge perturbation as
\begin{align}
    h^{(q)}_{\tau\tau}
    &=
    -N^2(x)\chi^2H(x)P_2(\cos\theta),
    \label{eq:hq_tautau_exactspin}
    \\
    h^{(q)}_{\tau\varphi}
    &=
    0,
    \label{eq:hq_tauphi_exactspin}
    \\
    h^{(q)}_{\varphi\varphi}
    &=
    x^2\sin^2\theta\,\chi^2K(x)P_2(\cos\theta),
    \label{eq:hq_phiphi_exactspin}
\end{align}
where
\begin{equation}
    N^2(x)
    =
    1-\frac{2}{x}-\frac{\lambda x^2}{3}.
    \label{eq:static_lapse_exactspin_source}
\end{equation}
The functions $H$ and $K$ are dimensionless radial Hartle--Thorne response functions in the leading even-parity \(\ell=2\) sector.  For the chosen kinetic profile they follow from the reduced radial Einstein equations in Appendix~\ref{app:nzflux_technical}.  The explicit factor $\chi^2$ fixes the slow-rotation order and makes the perturbation vanish in the non-rotating limit.  The ansatz contains only the leading even-parity $\ell=2$ terms. A full non-zero-charge Hartle--Thorne solution can also contain monopole terms and a charge correction to frame dragging, while a full exact-spin solution would contain higher powers of \(\chi\) and additional angular harmonics.

Substituting Eqs.~\eqref{eq:hq_tautau_exactspin}--\eqref{eq:hq_phiphi_exactspin} into the zero-charge Kerr--(A)dS Killing part gives
\begin{align}
    g_{\tau\tau}
    &=
    \bar g_{\tau\tau}
    -
    \varepsilon_q^2\chi^2N^2(x)H(x)P_2(\cos\theta)
    +
    \mathcal O(\varepsilon_q^2\chi^4,\varepsilon_q^4\chi^2),
    \label{eq:charged_gtautau_exactspin}
    \\
    g_{\tau\varphi}
    &=
    \bar g_{\tau\varphi}
    +
    \mathcal O(\varepsilon_q^2\chi),
    \label{eq:charged_gtauphi_exactspin}
    \\
    g_{\varphi\varphi}
    &=
    \bar g_{\varphi\varphi}
    +
    \varepsilon_q^2\chi^2x^2\sin^2\theta\,K(x)P_2(\cos\theta)
    +
    \mathcal O(\varepsilon_q^2\chi^4,\varepsilon_q^4\chi^2).
    \label{eq:charged_gphiphi_exactspin}
\end{align}
We use Eq.~\eqref{eq:charged_gphiphi_exactspin} as the circular-CTC diagnostic.  This is justified for the direct circular test because \(g_{\tau\varphi}\) does not enter the sign of the closed axial orbit.  We then compute the determinant and use it as a Killing-sector time-function check.

\section{Chronology diagnostics in rotating k-essence black holes}
\label{sec:kessenceCTCs}

A solution passes the chronology test if the axial Killing orbits remain spacelike in the exterior spacetime and the Killing-sector condition $\Delta_{\rm K}<0$ is maintained.  The first condition is the circular-CTC test; the second is a local consistency check.
We count a zero crossing of $g_{\varphi\varphi}$ as physical only when the perturbative control criteria in Appendix~\ref{app:numerical_implementation} are satisfied.

The zero-charge branch is exactly Kerr--(A)dS, with effective cosmological constant $\Lambda_{\rm eff}$ defined in Eq.~\eqref{eq:kessence_zerocharge_Einsteineqs}.  Equations~\eqref{eq:kerrads_det_identity} and \eqref{eq:kerrads_gphiphi_positive_identity_dimless} show that, after imposing the usual domain restrictions, the zero-charge branch has no exterior circular CTCs. The determinant $\Delta_{\rm K}$ vanishes only at horizons or on the rotation axis.
To study the AdS region, we choose $s_\alpha=-1$, $B<0$.  The analysis then splits into two geometrically distinct cases.  The usual Kerr--AdS branch has
$\Xi>0$ and therefore $B<-
    C_\Lambda\chi^2/12$,
which is a constraint on the geometry rather than on k-essence. The continuation with $\Xi=0$ or $\Xi<0$ is the nonstandard AdS extension analyzed in Sec.~\ref{sec:nonstandard_ads_zero_charge}.  For $\Xi<0$ the equatorial root satisfies Eq.~\eqref{eq:path_ads_equatorial_root}, and the circular-CTC region extends outward to the AdS boundary. We keep these cases for illustration, but label them as nonstandard-domain points.

\subsection{Nonzero-charge sector}

We analyze the nonzero-charge sector using Eq.~\eqref{eq:charged_gtautau_exactspin} together with $\Delta_K$.  The functions $H$ and $K$ come from the leading Hartle--Thorne radial response problem in Appendix~\ref{app:nzflux_technical}.  We solve the AdS-like branch $\lambda<0$ as an exterior problem outside the black-hole horizon, as appropriate for the Hartle--Thorne expansion. By contrast, we solve the dS-like branch $0<\lambda$ as a finite static-patch problem between the black-hole and cosmological horizons, with an additional regularity condition at the internal singular point of the radial response system.  In both cases the external current and residual stress are those described in Appendix~\ref{app:exact_source_closure}.

For fixed $(\chi,\theta,\varepsilon_q,C_X)$ and a specified branch, we define $D(\lambda,\chi)$ as the allowed radial domain.  For $\lambda<0$, $D$ is the domain outside the outer rotating black-hole horizon, with $\Xi>0$ and $\Delta_\theta>0$.  For $\lambda>0$, $D$ is the rotating static patch between the outer black-hole horizon and the cosmological horizon, again subject to $\Xi>0$ and $\Delta_\theta>0$. We can then define
\begin{align}
    \mathcal G(\lambda,\chi)
    \equiv
    \min_{x\in D(\lambda,\chi)}
    g_{\varphi\varphi}(x,\theta;\lambda,\chi,\varepsilon_q,C_X),
    \label{eq:G_min_exactspin}
    \quad
    \mathcal T(\lambda,\chi)
    \equiv
    \min_{x\in D(\lambda,\chi)}
    \big[-\Delta_{\rm K}(x,\theta;\lambda,\chi,\varepsilon_q,C_X)\big],
\end{align}
where chronology safety requires $\mathcal G>0$ and $\mathcal T>0$. 
We start from the Kerr limit, $\lambda=0$, and vary $\lambda$ along the chosen branch.  At each point we compute the minimum of $g_{\varphi\varphi}$ in the allowed radial domain and the minimum of $-\Delta_{\rm K}$.  Moving away from $\lambda=0$, we flag the first value of $\lambda$ at which either quantity reaches zero.
The interpretation depends on which quantity vanishes: if
$
\mathcal G(\lambda,\chi)=0,
$
then $g_{\varphi\varphi}=0$ somewhere in the exterior domain, and for nearby parameter values the closed azimuthal orbits can become timelike, which is the onset of circular CTCs.  If instead
$
\mathcal T(\lambda,\chi)=0,
$
then $\Delta_{\rm K}=0$ somewhere in the domain.  This signals the loss of the local Killing-sector time-function diagnostic, but it does not by itself guarantee the existence of a closed timelike curve.
We translate the effective variables back to model parameters using
\begin{equation}
    C_\Lambda\equiv \kappa\Mstar^4(GM)^2,
    \qquad
    \lambda
    =
    \Lambda_{\rm eff}(GM)^2
    =
    \frac{C_\Lambda}{4B}
    =
    \frac{C_\Lambda}{4}\frac{\alpha^2}{\tilde\beta} .
    \label{eq:lambda_normalized_alpha_beta_bound}
\end{equation}
Let $\mathcal I_{\rm CTC}$ denote the set of $\lambda$ values for which the first crossing is a controlled zero of $g_{\varphi\varphi}$, rather than a $\Delta_{\rm K}$ crossing or an artifact of the truncation.  Avoiding controlled circular CTCs with the source included requires
\begin{equation}
    \frac{C_\Lambda}{4B}
    \notin
    \mathcal I_{\rm CTC},
    \qquad
    \alpha\tilde\beta>0,
    \qquad
    \frac{\alpha^2}{2|\tilde\beta|}<X_{\rm max},
    \label{eq:ctc_avoidance_general_bound}
\end{equation}
where the last two inequalities are the static constant-$X$ branch and low-gradient conditions.
For example, we can take an AdS interval $\lambda\in[-L_2,-L_1]$ with $0<L_1<L_2$, which leads to the bounds
\begin{equation}
    -\frac{C_\Lambda}{4L_1}
    <
    B
    <
    -\frac{C_\Lambda}{4L_2},
    \qquad
    B<-\frac{C_\Lambda\chi^2}{12},
    \label{eq:ads_controlled_ctc_bound}
\end{equation}
where the second inequality is just $\Xi>0$.  Equivalently, the dS exclusion is $4\lambda_1/C_\Lambda<\alpha^2/\tilde\beta<4\lambda_2/C_\Lambda$, while the AdS exclusion is $4L_1/C_\Lambda<\alpha^2/|\tilde\beta|<4L_2/C_\Lambda$ together with the standard Kerr--AdS domain restriction. 

For the dS-like branch, $s_\alpha=+1$ and $B>0$.  There is no $\Xi>0$ pathology analogous to the AdS-like $\Xi=0$ boundary for positive $\lambda$ in the usual static patch. This branch exists only where the rotating black-hole and cosmological horizons bound a stationary radial region. We include it only when the finite static-patch boundary-value problem is solved and the circular root satisfies the control criterion.

In the numerical solution of the nonzero-charge case, we consider $g_{\varphi\varphi}<0$ a circular CTC candidate, check $\Delta_K$ and call it a controlled crossing iff the charged correction remains within EFT control over the full domain.

The dS-like zero-charge branch has $s_\alpha=+1$ and $B>0$ and is evaluated in the rotating static patch.  The usual AdS zero-charge branch has $s_\alpha=-1$, $B<0$, and $\Xi>0$.  Both zero-charge branches are safe against exterior circular CTCs in their usual domains.  The nonstandard AdS continuation, with $\Xi=0$ or $\Xi<0$, is kept only as a domain diagnostic; for $\Xi<0$ the circular-CTC region extends asymptotically and is not counted as a controlled sourced effect.  The charged branch uses the usual dS or AdS radial domains together with the leading $\chi^2\varepsilon_q^2$ response, and possible circular roots are kept only when the control criteria are satisfied.  The echo model is separate: it requires an imposed wall or excision surface with $x_+<x_{\rm wall}<x_{\rm peak}$ and is not inferred from the CTC diagnostic alone.
Table~\ref{tab:numerical_control_summary} gives the main numerical result, and
\begin{figure}[t]
    \centering
\includegraphics[width=0.98\linewidth]{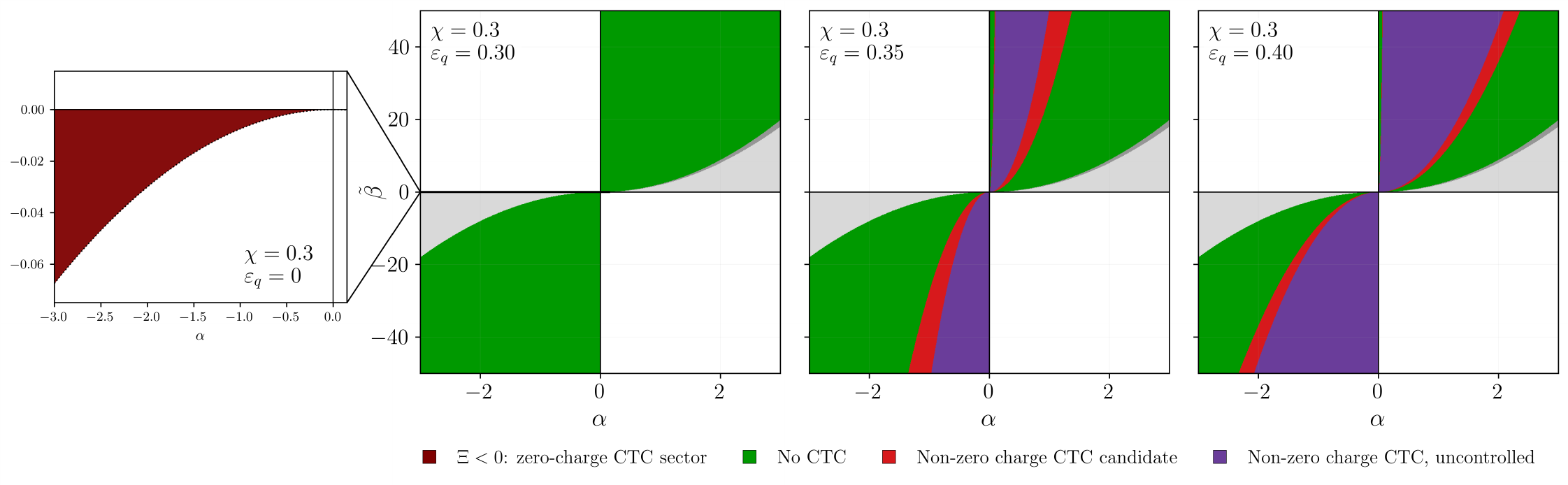}
    \caption{Circular-CTC diagnostic in the $(\alpha,\tilde\beta)$ plane for fixed spin $\chi=0.3$ and three values of the source amplitude, $\varepsilon_q=\{0.30,0.35,0.40\}$.  The main panels show the nonzero-charge case in the usual Kerr--(A)dS domains, keeping the leading Hartle--Thorne circular response at $\mathcal O(\chi^2\varepsilon_q^2)$.  Green regions have $g_{\varphi\varphi}>0$.  Red points are nonzero-charge CTC candidates that pass the EFT control cut, while purple points are possible sign changes that fail the control test and are not considered under EFT control. The plot uses the order-one cut $c_{\rm max}=1.1$.  White regions lie off the static constant-$X$ branch, and grey regions fail the imposed low-gradient cutoff. The inset magnifies the narrow region near $\tilde\beta=0$ where $\Xi<0$ at zero charge. The dark-red band is the formal zero-charge Kerr--AdS CTC sector,
    $-C_\Lambda\chi^2\alpha^2/12<\tilde\beta<0$, shown on the allowed branch with $\alpha<0$. This sector already contains CTCs at zero charge.}
    \label{fig:alpha_beta_ctc_scan}
\end{figure}
\begin{table}[t]
\centering
\def\arraystretch{1.0}
\begin{tabular}{cccccc}
\toprule
$\chi$ & $\varepsilon_q$ & $c_{\rm max}$ & AdS interval & dS interval & smallest required $c_{\rm max}$ \\
\midrule
0.30 & 0.30 & 0.3 & -- & -- & -- \\
0.30 & 0.35 & 0.3 & -- & -- & -- \\
0.30 & 0.40 & 0.3 & -- & -- & -- \\
0.30 & 0.30 & 1.1 & -- & -- & -- \\
0.30 & 0.35 & 1.1 & -- & $0.0050\lesssim\lambda\lesssim0.0095$ & $1.016$ \\
0.30 & 0.40 & 1.1 & -- & $0.0219\lesssim\lambda\lesssim0.0279$ & $1.014$ \\
\bottomrule
\end{tabular}%
\caption{Main summary corresponding to the nonzero-charge panels of Fig.~\ref{fig:alpha_beta_ctc_scan}.  The strict perturbative criterion $c_{\rm max}=0.3$ gives no controlled circular CTCs for the displayed charge amplitudes.  With the order-one cut $c_{\rm max}=1.1$, the $\varepsilon_q=0.35$ and $\varepsilon_q=0.40$ cases contain possible dS-side sign changes.}
\label{tab:numerical_control_summary}
\end{table}
Figure~\ref{fig:alpha_beta_ctc_scan} classifies each point in the parameter plane. We first map the point to $\lambda=C_\Lambda\alpha^2/(4\tilde\beta)$ and then apply the branch, cutoff, domain, boundary-value-problem, and CTC tests.  White points lie off the static constant-$X$ branch, $\alpha\tilde\beta<0$. Grey points fail the low-gradient condition $\alpha^2/(2|\tilde\beta|)<X_{\rm max}$.  Green points have no exterior zero crossing of $g_{\varphi\varphi}$ and are therefore chronologically safe. Red points are nonzero-charge regions where $g_{\varphi\varphi}<0$ acccording to the control cut which in Figure~\ref{fig:alpha_beta_ctc_scan} is $c_{\rm max}=1.1$, allowing for unity-size perturbations. Purple points also show possible crossings, but fail the control test and lie beyond EFT validity.  The dark-red inset shows the {\it zero}-charge $\Xi<0$ Kerr--AdS sector.

With the strict control cuts $\mathcal C_{\rm amp}<0.3$ and $\mathcal C_{\varphi\varphi}<0.3$, we find no controlled nonzero-charge circular CTC for $\chi=0.3$. This follows because, for a zero-charge background obeying the usual domain restrictions, $g_{\varphi\varphi}^{(0)}>0$, so a zero of $g_{\varphi\varphi}^{(0)}+\delta g_{\varphi\varphi}$ requires $|\delta g_{\varphi\varphi}/g_{\varphi\varphi}^{(0)}|\gtrsim1$ near the crossing. The red points therefore show what happens when the control criterion is relaxed to order unity. In the normalization $C_\Lambda=1$, the $\varepsilon_q=0.35$ case implies
$
    26.3\lesssim B\lesssim 50.0,
$
and increasing the charge slightly to $\varepsilon_q=0.40$ gives
$
    8.96\lesssim B\lesssim 11.4 .
$
For $\varepsilon_q=0.30$, the metric never reaches $g_{\varphi\varphi}=0$ in the usual domain, even with the looser value $c_{\rm max}=1.1$.  With the stricter perturbative value $c_{\rm max}=0.3$, the same conclusion holds for all three displayed charge amplitudes: there is no controlled point at which the closed azimuthal orbits become timelike. This gives no new bound on $\alpha$ or $\beta$. However, it shows that geometric causality bounds require more than a slow-spin Hartle-Thorne expansion and perturbative charge.  

No robust exclusion appears at this order. The zero-charge Kerr--(A)dS branch is chronology-safe after the usual domain restrictions are imposed. The nonzero-charge branch can produce zero crossings of $g_{\varphi\varphi}$, and an order-one control criterion makes some of them visible as red regions in Fig.~\ref{fig:alpha_beta_ctc_scan}. The zero-charge Kerr--(A)dS space is exact in $\chi$, but the nonzero-charge crossings are obtained using the leading Hartle--Thorne response at $\mathcal O(\chi^2\varepsilon_q^2)$. Chronology-based bounds on rotating solutions or model parameters require going beyond the leading Hartle--Thorne approximation and including higher-order backreaction from the charge $q$.

\section{Echoes as probes of geometric causality}
\label{sec:echoes}

Gravitational-wave echoes are delayed, ringdown-like pulses that can arise when radiation is partially trapped between an inner reflective surface and the angular-momentum barrier near the photon sphere~\cite{Cardoso:2016rao,Cardoso:2016oxy,Mark:2017dnq,Cardoso:2019rvt}. The existence of echoes therefore requires a semi-reflective surface at or near the horizon, usually ascribed as a quantum gravity effect. 

Suppose instead that the chronology-violation surface is replaced by some short-distance completion with a partially reflective boundary. This boundary condition is an extra assumption: a CTC root alone does not impose reflection nor predicts echoes.

We describe the scattering problem with a probe scalar on the exact zero-charge Kerr--((A))dS geometry.  The k-essence fluctuation is not the propagating field on the $P_X=0$ branch.  Instead we introduce a probe scalar $\Phi$ satisfying
\[
\Box_{g^{(0)}}\Phi=0,
\]
where $g^{(0)}_{\mu\nu}$ is the Kerr--((A))dS metric in Eq.~\eqref{eq:Kerr-AdS_metric}.  With the separated form
$
\Phi
=
e^{-i\omega t}e^{im\varphi}
S_{\ell m}(\theta)R_{\ell m\omega}(r),
$
the radial equation can be written as
\[
\frac{d}{dr}
\left(
\Delta_r\frac{dR_{\ell m\omega}}{dr}
\right)
+
\left[
\frac{K_r^2}{\Delta_r}
-
\lambda_{\ell m}
\right]
R_{\ell m\omega}
=
0,
\quad
K_r=(r^2+a^2)\omega-a\Xi m ,
\]
where $\lambda_{\ell m}$ is fixed by the angular equation. After introducing the tortoise coordinate $r_*$ as
\begin{equation}
    \frac{dr_*}{dr}
=
\frac{r^2+a^2}{\Delta_r(r)}
\label{eq:tortoise_main_echo}
\end{equation}

and redefining the radial field to remove the first derivative, the equation takes the schematic Schrödinger form
\[
\frac{d^2\Psi_{\ell m\omega}}{dr_*^2}
+
\left[
\omega^2
-
V_{\ell m}(r;\omega)
\right]
\Psi_{\ell m\omega}
=
0 .
\]
The ordinary ringdown follows from this equation with the usual QNM boundary conditions: ingoing at the black-hole horizon and the appropriate outgoing, regular, or normalizable condition at the outer boundary, depending on whether the exterior is asymptotically flat, dS, or AdS.  In the echo setup we change the inner boundary condition.  Instead of imposing perfect absorption at the horizon, we impose a partially reflective condition at
$r_{\rm wall}=GM x_{\rm wall}$. Near the wall we write
\begin{equation}
\Psi_\omega(y)=A_{\rm in}e^{-i\omega y}+A_{\rm out}e^{i\omega y},
\quad
y=r_*-r_{*,{\rm wall}},
\quad
{\cal R}_{\rm wall}(\omega)=\frac{A_{\rm out}}{A_{\rm in}} .
\label{eq:wall_reflectivity_bc}
\end{equation}
Equivalently, a local Robin membrane condition
$
\partial_{r_*}\Psi_\omega=Z(\omega)\Psi_\omega
$
at the wall gives
$
{\cal R}_{\rm wall}(\omega)=(i\omega+Z)/(i\omega-Z).
$
Thus the primary ringdown comes from the same probe-scalar wave equation as the ordinary QNM problem, while the echoes arise after replacing the absorbing inner boundary by a partially reflective one.  The delayed-copy formula below is a narrow-band time-domain representation of this modified scattering problem.

Let $x_+$ be the outer black-hole horizon and let $x_{\rm peak}$ be the maximum of the relevant perturbation potential, approximated by the photon-ring barrier in the eikonal regime.  A valid exterior cavity must satisfy
\begin{equation}
    x_+<x_{\rm wall}<x_{\rm peak},
    \qquad
    \Delta_x>0 \quad (x_{\rm wall}<x<x_{\rm peak}),
    \qquad
    \mathcal N_{\varphi}(x,\theta)\ge0 \quad (x_{\rm wall}\le x\le x_{\rm peak}) .
    \label{eq:echo_cavity_conditions_revised}
\end{equation}
For the equatorial $\Xi<0$ zero-charge branch, the last condition requires the proposed cavity not to intersect the outward CTC region.  If $x_{\rm CTC}<x_{\rm peak}$, setting $x_{\rm wall}=x_{\rm CTC}$ would put the immediately exterior interval in the CTC region and is not a standard echo cavity.
From the Kerr-(A)dS tortoise coordinate $r_*$ and
using $r=GMx$, $a=GM\chi$, and $\Delta_r=(GM)^2\Delta_x$, the round-trip echo delay is
\begin{equation}
    \frac{\Delta t_{\rm echo}}{GM}
    =
    2
    \int_{x_{\rm wall}}^{x_{\rm peak}}
    \frac{x^2+\chi^2}{
        (x^2+\chi^2)
        \left(
            1-\frac{\lambda x^2}{3}
        \right)
        -2x
    }
    \,dx .
    \label{eq:echo_delay_dimensionless_main}
\end{equation}
With this definition of the echo time delay, we can define the tortoise size of the echo cavity as $L_{\rm cav}=r_*(r_{\rm peak})-r_*(r_{\rm wall})=\Delta t_{\rm echo}/2$.
A wave that completes one round trip in the cavity is multiplied by
$
{\cal R}_{\rm wall}(\omega)
{\cal R}_{\rm ph}(\omega)
e^{2i\omega L_{\rm cav}},
$
where ${\cal R}_{\rm wall}$ is the wall reflectivity and ${\cal R}_{\rm ph}$ is the reflection coefficient of the photon-sphere barrier for a wave incident from inside the cavity. Resonant cavity modes occur when the wave reproduces itself after one round trip,
$
{\cal R}_{\rm wall}(\omega)
{\cal R}_{\rm ph}(\omega)
e^{2i\omega L_{\rm cav}}=1 .
$
Equivalently,
$
1-
{\cal R}_{\rm wall}(\omega)
{\cal R}_{\rm ph}(\omega)
e^{2i\omega L_{\rm cav}}
=0 ,
$
which is the pole condition. The real part of the resulting complex frequency fixes the resonance spacing, while the imaginary part gives the damping due to imperfect reflection and leakage out of the cavity.
Writing ${\cal R}_{\rm wall}{\cal R}_{\rm ph}=|{\cal R}_{\rm wall}{\cal R}_{\rm ph}|e^{i\Phi_R}$ over a narrow band where the phase shift $\Phi_R$ is taken as constant gives
\begin{equation}
    \omega_n
    \simeq
    \frac{\pi n-\Phi_R/2}{L_{\rm cav}}
    -
    \frac{i}{2L_{\rm cav}}
    \ln\left[
        \frac{1}{|{\cal R}_{\rm wall}{\cal R}_{\rm ph}|}
    \right],
    \qquad
    \Delta\omega\simeq\frac{\pi}{L_{\rm cav}}
    =
    \frac{2\pi}{\Delta t_{\rm echo}} .
    \label{eq:conditional_echo_modes}
\end{equation}
In the time domain, the waveform can be written as a sum of repeated copies of the ordinary ringdown.  The first term, \(n=0\), is the original GR-like ringdown \(h_0(t)\).  The later terms, \(n\ge1\), are echoes: each one is delayed by an additional time \(\Delta t_{\rm echo}\), multiplied by an amplitude factor \(R\), and shifted by a phase \(\phi_R\). 

The total ringdown signal can then be written as
\begin{equation}
    h(t)=\sum_{n=0}^{\infty}R^n e^{in\phi_R}h_0(t-n\Delta t_{\rm echo}),
    \label{eq:echo_delayed_copy}
\end{equation}
which gives the transfer function in Fourier space as
\begin{equation}
    \tilde h(\omega)
    =
    \frac{\tilde h_0(\omega)}{1-Re^{i(\phi_R-\omega\Delta t_{\rm echo})}},
    \qquad
    |\tilde h(\omega)|^2
    =
    \frac{|\tilde h_0(\omega)|^2}{1+R^2-2R\cos(\phi_R-\omega\Delta t_{\rm echo})} .
\label{eq:echo_transfer_delayed_copy}
\end{equation}
For the representative time-domain plot we model the dominant probe-scalar QNM by a causal damped sinusoid beginning at \(t=t_0\).  This is the $h_0(t)$ entering Eq.~\eqref{eq:echo_delayed_copy}.  The geometry fixes the QNM problem and the echo delay, while the effective wall reflectivity fixes the relative amplitude and phase of the echoes.

\begin{figure}
    \centering
    \includegraphics[width=0.9\linewidth]{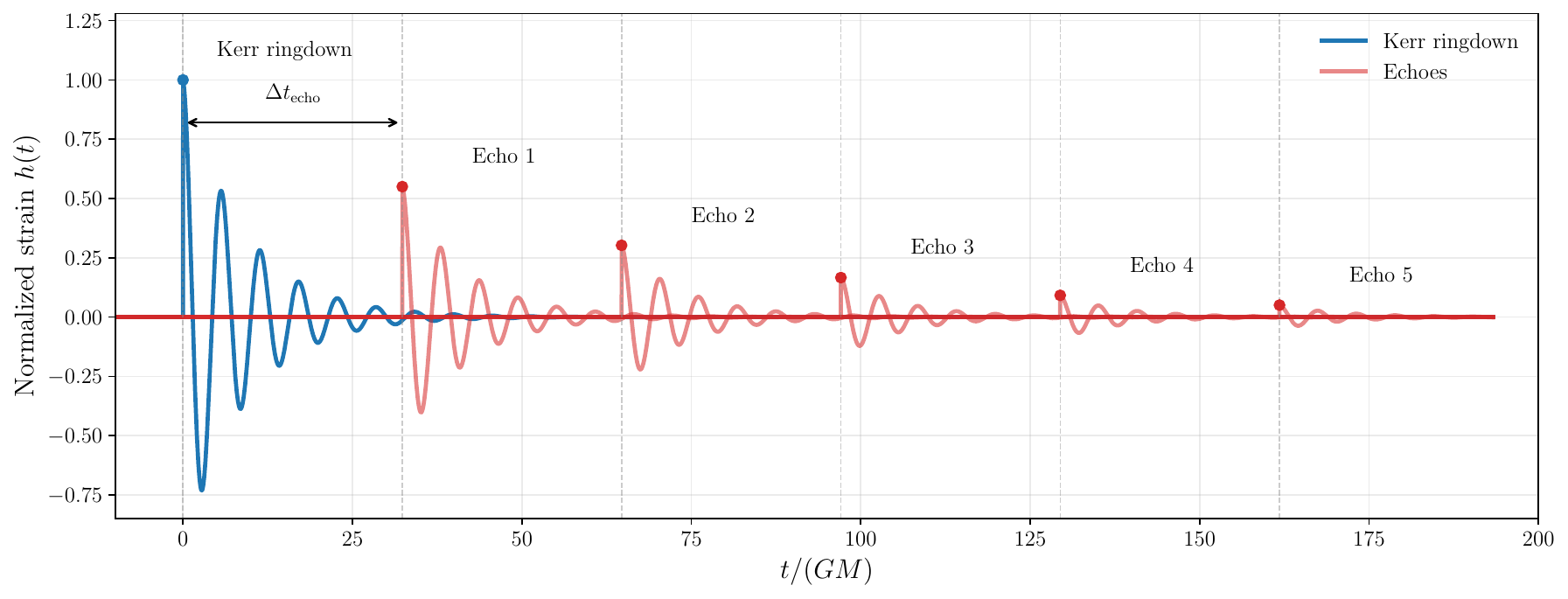}
    \caption{Echo train generated by the delayed-copy model of Eq.~\eqref{eq:echo_delayed_copy}.  The first packet is the ordinary Kerr-like ringdown $h_0(t)$, while the later packets are echoes delayed by integer multiples of $\Delta t_{\rm echo}$.  The spacing is fixed by the imposed wall-to-barrier cavity through Eq.~\eqref{eq:echo_delay_dimensionless_main}.  The decrease of the isolated echo envelopes follows $A_n/A_0=R^n$ for a constant effective round-trip amplitude $0<R<1$.}
    \label{fig:echo-plot}
\end{figure}

The term \(n=0\) in Eq.~\eqref{eq:echo_delayed_copy} is therefore
the GR ringdown itself, while the terms \(n\ge1\) are echoes delayed by
integer multiples of \(\Delta t_{\rm echo}\).  In the simplest narrow-band
model \(R\) is treated as a constant, \(0<R<1\). The echo amplitudes then scale as
    $A_n/A_0 = R^n,
    \,
    A_{n+1}/A_n = R$,
up to phase-dependent interference when the delayed copies overlap.
Figure~\ref{fig:echo-plot} shows the time-domain content of this echo model. The first visible packet is the ordinary
GR-like ringdown \(h_0(t)\).  In the delayed-copy convention of
Eq.~\eqref{eq:echo_delayed_copy}, this is the \(n=0\) term in the sum and it is not itself an echo.  The subsequent packets are the \(n\ge1\)
terms,
\begin{equation}
    h_n(t)
    =
    R^n e^{in\phi_R}
    h_0(t-n\Delta t_{\rm echo}) ,
    \quad n\ge1 ,
\end{equation}
so their arrival times are shifted by integer multiples of the round-trip
cavity time.  Thus the horizontal separation between successive echo
packets in Figure~\ref{fig:echo-plot} is \(\Delta t_{\rm echo}\), up to the
small displacement between the start time of a damped sinusoid and the
location of its first local maximum.

The delay can also be decomposed into a reference travel time and a Shapiro-time-delay calculation. The echo spacing is twice the one-way travel time between the wall and the photon-sphere barrier, measured in the tortoise coordinate.  We can therefore separate the travel time into a reference propagation time and a curvature-induced delay. Approximately, we can take
$
\Delta t_{\rm echo}/2GM
\simeq
(x_{\rm peak}-x_{\rm wall})
+
\Delta t_{\rm Shapiro}/2GM, $ where the Shapiro time delay can be taken as $
\Delta t_{\rm Shapiro}
=
2GM\int_{x_{\rm wall}}^{x_{\rm peak}}
\left[
(x^2+\chi^2)/\Delta_x
-1
\right]dx .
$
In an asymptotically dS or AdS setup, one should instead subtract the corresponding no-black-hole reference travel time, but the interpretation is unchanged.  The echo delay measures the extra propagation time accumulated inside the curved cavity.  This connects the echo construction to the recent use of time delays as causality diagnostics in gravitational EFTs: a controlled EFT should not generate a resolvable time advance within its regime of validity~\cite{deRham:2021bll,Chen:2021bvg,Melville:2024zjq}.  In the present setup the relevant observable is not a scattering time delay from infinity to infinity, but a cavity time delay between two imposed radial surfaces.  Once a reflective wall is imposed, the measured echo spacing probes whether the effective geometry increases or decreases the propagation time relative to an appropriate reference background.  Moving the wall closer to the horizon increases the delay because the tortoise coordinate develops the usual logarithmic enhancement near $x_+$.

The value of \(\Delta t_{\rm echo}\) used in the plot is fixed by the
geometry of the cavity.  The wall and barrier must first satisfy the
ordering and regularity conditions in
Eq.~\eqref{eq:echo_cavity_conditions_revised}. Once the cavity is
specified, we compute the propagation time from the Kerr--(A)dS tortoise
coordinate in Eq.~\eqref{eq:tortoise_main_echo}.  In dimensionless form this
gives Eq.~\eqref{eq:echo_delay_dimensionless_main}, and changing \(x_{\rm wall}\), \(x_{\rm peak}\), \(\chi\), or \(\lambda\)
changes the spacing of echoes through the cavity length.  Moving the wall closer to
the horizon typically increases the tortoise separation and therefore
increases \(\Delta t_{\rm echo}\); moving the wall toward the photon-sphere
barrier shortens the cavity and decreases the echo spacing.
A longer cavity produces more widely separated echoes
in time and a more finely spaced resonance comb in frequency.  Conversely, a
shorter cavity produces more closely spaced echoes and a larger
\(\Delta\omega\). 
Thus
Fig.~\ref{fig:echo-plot} shows the consequence of treating the chronology violation surface as partially reflected in the manner of echo phenomenology: the \(n=0\) GR ringdown is followed by delayed
echoes separated by \(\Delta t_{\rm echo}\), with isolated amplitudes
decreasing as \(R^n\)\footnote{A rotating reflective cavity must also satisfy the usual superradiant and ergoregion-instability constraints.  In the superradiant band, the net round-trip gain must remain below unity after including wall reflectivity, barrier amplification, absorption, and any UV dissipation~\cite{Press:1972zz,Maggio:2018ivz}.}.

A viable echo cavity requires that a reflective or excision boundary is imposed and the interval between the wall and the photon-sphere barrier satisfies Eq.~\eqref{eq:echo_cavity_conditions_revised}, the model produces echoes with the delay fixed by Eq.~\eqref{eq:echo_delay_dimensionless_main} and the frequency spacing fixed by Eq.~\eqref{eq:conditional_echo_modes}.  If the root lies outside the barrier, or if the wall-to-barrier interval intersects a CTC region, the root does not define a healthy echo cavity.  A CTC root alone is therefore not an echo prediction, and the parameter-space in Fig.~\ref{fig:alpha_beta_ctc_scan} is not an echo plot.  Even where the looser diagnostic labels a circular zero as passing the display cut, an echo prediction still requires an additional reflective boundary and the cavity ordering in Eq.~\eqref{eq:echo_cavity_conditions_revised}. Including linear spin-2 perturbations in the ringdown and not rely on a probe scalar, we would need to solve the corresponding Teukolsky or Sasaki-Nakamura problem. For Kerr--((A))dS and for the source-supported backgrounds considered here, the appropriate ringdown equation and boundary conditions must be specified separately. We leave this for future work.

\section{Discussion \& Outlook}
\label{sec:disc}

We have used closed timelike curves as a geometric causality diagnostic in scalar-tensor gravity. For this purpose, we study the azimuthal metric component $g_{\varphi\varphi}$ supplemented by the Killing-sector determinant criterion of Sec.~\ref{sec:CTCcriterion}.  In the zero-charge branch, we use the full Kerr--(A)dS Killing part and work to all orders in $a/(GM)$. By contrast, in the non-zero charge case we use a Hartle--Thorne slow-rotation calculation and we keep only the leading quadrupolar response at $\mathcal O(\chi^2\varepsilon_q^2)$.

In the k-essence model, the non-canonical kinetic structure introduces a branch dependence controlled by $s_\alpha$ and $B$.  In the zero-charge branch, the scalar stress tensor is vacuum-energy-like and the geometry is exactly Kerr--(A)dS.  After imposing $\Xi>0$, $\Delta_\theta>0$, and the appropriate exterior or static-patch radial domain, this branch has no exterior circular CTCs.  The nonstandard AdS extension with $\Xi\le0$ can be analyzed explicitly, but it is not part of the usual Kerr--AdS problem: $\Xi=0$ is a boundary degeneracy and $\Xi<0$ gives an asymptotic, outward-extending circular-CTC region.
With the diagnostic used in this paper, only the sector with the external source can produce possible circular close d timelike curves, since the quadrupolar contribution (the first order which can generate CTCs) vanish in the sourceless limit.  The external shift current supports $\mathcal X_{2,2}\neq0$ and hence nonzero $\Pi_2$ and $S_K$, whereas in the sourceless limit we have that $\mathcal X_{2,2}=S_K=\Pi_2=0$ which removes all $g_{\varphi\varphi}$ regions. In the non-zero charge case, we combine the exact zero-charge Kerr--(A)dS space with the leading Hartle--Thorne terms in $g_{\varphi\varphi}$, namely the $\mathcal O(\chi^2\varepsilon_q^2)$ quadrupolar sector.  We solve the AdS-like branch as an exterior problem, while the dS-like branch requires a finite static-patch boundary-value problem with regular matching at an internal singular point of the radial response system.

We also considered imposing a partially reflective boundary at the chronology-violation surface. In this case, the echo spacing $\Delta t_{\rm echo}$ is a derived property from the model parameters, whereas the phase shift and amplitude drop are consequences of the reflective wall which is input by hand.  If a future source completion or parameter scan produces a perturbatively controlled circular root outside the horizon and inside the photon-sphere barrier, a separate ultraviolet boundary condition could turn that surface into a partially reflective wall. The exact Kerr--(A)dS tortoise coordinate would then fix the leading cavity length.
Current LIGO--Virgo--KAGRA searches have uncovered no statistically significant evidence for gravitational-wave echoes. Recent model-independent analyses, including high-ringdown-SNR events from the O4 era, place 90\% upper limits on late-time echo-like strain amplitudes, with representative time-domain bounds around $10^{-24}$ for the average long-lived post-merger signal~\cite{Wu:2025enn}. The first part of the fourth observing run, O4a, has nevertheless enlarged the available event sample substantially through GWTC-4.0, including new high-SNR binary-black-hole mergers that are especially useful for ringdown and post-ringdown tests~\cite{LIGOScientific:2025slb}. Earlier claims of echo-like residuals in individual post-merger data sets remain at low statistical significance and are strongly analysis-dependent~\cite{Abedi:2016hgu,Conklin:2017lwb,Tsang:2023gxp}. Echoes are intrinsically challenging to detect because they are expected to appear as quasi-periodic, slowly damped post-ringdown signals whose amplitudes are typically at most a small fraction of the main merger-ringdown amplitude, often $\lesssim 10\% -20\%$ in phenomenological templates. Robust searches therefore require waveform models or excess-power methods that account for spin-dependent phase shifts, frequency-dependent reflectivity, damping, and possible mode mixing~\cite{Mark:2017dnq,Wang:2019rcf}. Next-generation detectors such as the Einstein Telescope, Cosmic Explorer, and LISA should improve the sensitivity to delayed post-merger structure and may probe echo time delays, angular dependence, and mode content in the theoretically interesting regime~\cite{Maggio:2020jml,Barausse:2020rsu}. Such observations would provide a direct test of near-horizon structure, quantum-gravity corrections, and the causal properties of modified-gravity EFTs.

CTC avoidance provides a practical consistency test for modified-gravity backgrounds.  It does not replace positivity, hyperbolicity, or Shapiro-delay bounds, although the Shapiro delay can be related to CTCs thanks to its geometrical nature. Instead, CTCs add a global condition on the backreacted spacetime. We separate exact zero-charge chronology statements from nonzero-charge statements that still depend on slow rotation and a perturbative matter sector. Establishing a controlled bound will require a microscopic source completion and a higher-order, or fully nonlinear, rotating perturbation analysis.

\begin{acknowledgments}
The authors thank Pavel Petrov and Theodoros Nakas for useful discussions. B.H.L. is supported by the National Research Foundation of Korea (NRF) grant RS-2020-NR049598, RS-2026-25473640, RS-2026 25473640, RS-2024-00339204, and by the Overseas Visiting Fellow Program of Shanghai University. B.H.L. thanks the Asia Pacific Center for Theoretical Physics, Korea, for hospitality during the visit where part of this project was carried out. N.A.N. was supported by the Institute for Basic Science under project code IBS-R018-D3 and acknowledges support from PSL/Observatoire de Paris. S.T. thanks CQUeST for support during this work, as well as IIT-BHU for hospitality during the visit where a preliminary version of this paper was presented.
\end{acknowledgments}

\appendix
\begin{center}
{\bf APPENDICES}
\end{center}
\section{Rotating-cylinder backreaction}
\label{app:cyl_details}

This appendix gives the first-order scalar backreaction used in Sec.~\ref{sec:tipler}.  We work at \(\mathcal O(\varepsilon)\), with \(\varepsilon\) defined in Eq.~\eqref{eq:cylinder_scalar_ansatz}, on the rotating-cylinder background of Eq.~\eqref{eq:rotcyl}.
In the locally flat coordinate \(T=t-\omega\varphi\) we write
\begin{equation}
    ds^2= -[1+\varepsilon a(r)]dT^2+2\varepsilon b(r)dT\,d\varphi
    +[r^2+\varepsilon c(r)]d\varphi^2+dr^2+[1+\varepsilon s(r)]dz^2+\mathcal O(\varepsilon^2).
    \label{eq:metric-pert-app-T}
\end{equation}
Returning to the coordinate \(t=T+\omega\varphi\) gives the circular component in Eq.~\eqref{eq:gphiphi_full}. We use the scalar profile defined in Eq.~\eqref{eq:scalar_profile}.  The profile is externally regulated and is not a sourceless solution of the massless scalar equation.  With
\begin{equation}
    \Sigma(r)=\frac12\frac{r^2}{(r^2+r_0^2)^2},
    \label{eq:Sigma_app}
\end{equation}
the independent linearized equations can be written as
\begin{align}
    -\mathcal B &= \Sigma,
    &
    \omega\mathcal B-\frac12 b''+\frac{b'}{2r}&=-\omega\Sigma,
    &
    \frac{a'+s'}{2r}&=\Sigma,
    \label{eq:cyl_radial_eqs_compact}\\
    \frac12a''+\frac{a'}{2r}+\frac{c''}{2r^2}-\frac{c'}{r^3}+\frac{c}{r^4}&=-\Sigma,
    &
    \frac12r^2(a''+s'')+\omega b''-\frac{\omega b'}{r}-\omega^2\mathcal B&=-(r^2-\omega^2)\Sigma,
    \nonumber
\end{align}
where
\begin{equation}
    \mathcal B=\frac12s''+\frac{s'}{2r}+\frac{c''}{2r^2}-\frac{c'}{r^3}+\frac{c}{r^4}.
    \label{eq:B_app}
\end{equation}
The first two equations give \(b(r)=B_0+B_2r^2\).  We set the growing mode \(B_2=0\) and keep the constant mode \(B_0\).  Defining \(u=a+s\) and \(w=s-a\), regularity removes the logarithmic mode in \(w\), and with \(a(r_0)=s(r_0)=0\) we obtain
\begin{equation}
    a(r)=s(r)=\frac14\left[
        \ln\left(\frac{r^2+r_0^2}{2r_0^2}\right)+\frac{r_0^2}{r^2+r_0^2}-\frac12
    \right].
    \label{eq:a_s_sol_app}
\end{equation}
The remaining equation for \(c\) is
\begin{equation}
    c''-\frac2r c'+\frac{2}{r^2}c
    =-\frac{r^4(r^2+3r_0^2)}{(r^2+r_0^2)^3},
    \label{eq:c_ode_norm_app}
\end{equation}
whose regular solution, normalized by \(c(r_0)=0\), is
\begin{equation}
    c(r)=\frac12 r^2\ln\left(\frac{2r_0^2}{r^2+r_0^2}\right)
    +\frac{2+3\pi}{16}r^2
    -\frac34rr_0\arctan\left(\frac{r}{r_0}\right)
    -\frac{r^2r_0^2}{4(r^2+r_0^2)} .
    \label{eq:c_sol_norm_app}
\end{equation}
Substituting \(b=B_0\) into Eq.~\eqref{eq:gphiphi_full} gives the shift in the chronology root seen in Eq.~\eqref{eq:rctc_B0}; for arbitrary orientation one expands around \(|\omega|\).

\section{Source-supported radial system}
\label{app:nzflux_technical}
\label{vac_ell2_basis}

This appendix gives the radial system with the external source used in Secs.~\ref{sec:nzflux} and \ref{sec:kessenceCTCs}.  The zero-charge Kerr--(A)dS background is known exactly, but in the nonzero-charge radial system below we keep only the leading Hartle--Thorne even-parity sector.  This defines the quadrupolar deformation that enters $g_{\varphi\varphi}$ at $\mathcal O(\chi^2\varepsilon_q^2)$.

We write the radial response problem in terms of the dimensionless variables $x$ and $\lambda$ in Eq.~\eqref{eq:x_chi_lambda_def} and use the static lapse $N^2(x)$ defined in Eq.~\eqref{eq:static_lapse_exactspin_source}.  The roots of $N^2$ determine the singular points of the radial operator.  For $\lambda<0$, there is one positive black-hole root, denoted $x_b$.  For $0<\lambda<1/9$ there are two positive roots, the black-hole root $x_b$ and the cosmological root $x_c$. The functions $H(x)$ and $K(x)$ obey the differential equations
\begin{equation}
    \frac{d}{dx}
    \begin{pmatrix}
        H \\ K
    \end{pmatrix}
    =
    A(x;\lambda)
    \begin{pmatrix}
        H \\ K
    \end{pmatrix}
    +
    S(x;\lambda),
    \label{eq:HK_system_app}
\end{equation}
where
\begin{equation}
    A(x;\lambda)
    =
    \begin{pmatrix}
        c_{HH}(x;\lambda)-a_{KH}(x;\lambda)
        &
        -a_{KK}(x;\lambda)
        \\
        a_{KH}(x;\lambda)
        &
        a_{KK}(x;\lambda)
    \end{pmatrix},
    \qquad
    S(x;\lambda)
    =
    \begin{pmatrix}
        -S_K(x;\lambda)\\
        S_K(x;\lambda)
    \end{pmatrix}.
    \label{eq:app_A_S_matrix_nzflux}
\end{equation}
The coefficients are
\begin{align}
    a_{KK}(x;\lambda)
    &=
    \frac{6}{3-\lambda x^3},
    \label{eq:aKK_exact_source}
    \\
    a_{KH}(x;\lambda)
    &=
    \frac{2(6+6x+\lambda x^3)}
    {x(3-\lambda x^3)},
    \label{eq:aKH_exact_source}
    \\
    c_{HH}(x;\lambda)
    &=
    \frac{2(3-\lambda x^3)}
    {x(6-3x+\lambda x^3)}.
    \label{eq:cHH_exact_source}
\end{align}
The source is
\begin{equation}
    S_K(x;\lambda)
    =
    \frac{3x^2}
    {4(3-\lambda x^3)}
    \Pi_2(x;\lambda),
    \label{eq:app_SK_def_nzflux}
\end{equation}
with
\begin{equation}
    \Pi_2(x;\lambda)
    =
    -4\mathcal X_{2,2}(x)m_{\rm src}(x;\lambda),
    \label{eq:app_Pi2_def_nzflux}
\end{equation}
and the source-shape function
\begin{equation}
    m_{\rm src}(x;\lambda)
    =
    \frac{5-x}{x^4}
    +
    \frac{\lambda}{3x}.
    \label{eq:m_source_shape}
\end{equation}
Equations~\eqref{eq:HK_system_app}--\eqref{eq:m_source_shape} define the leading Hartle--Thorne radial response model used in the present paper. The zero-charge Kerr--(A)dS factors in the analysis are exact, but the charged response itself is kept only at $\mathcal O(\chi^2\varepsilon_q^2)$ for the circular component.

\subsection{External current and sourceless radial limit}

We define
\begin{equation}
    Z(r,\theta)
    \equiv
    \mathcal X_{2,0}(r)+\mathcal X_{2,2}(r)P_2(\cos\theta).
    \label{eq:app_Z_def}
\end{equation}
Here $Z$ is dimensionless, so the physical kinetic perturbation is $\delta X=\varepsilon_q^2\Mstar^4 Z$.  At the first order where $P_X\neq0$, the stationary radial background gives
\begin{equation}
    \nabla_\mu(P_X\nabla^\mu\phi)
    =
    \frac{2\tilde\beta\varepsilon_q^2}{r^2}
    \frac{d}{dr}
    \left[
        r^2N^2(r)\phi_0'(r)Z(r,\theta)
    \right]
    +
    \mathcal O(\varepsilon_q^4).
    \label{eq:app_scalar_divergence_Z}
\end{equation}
A convenient external current supporting this profile is purely radial at this order,
\begin{equation}
    J_{\rm ext}^{\mu}
    =
    2\tilde\beta\varepsilon_q^2
    N^2(r)\phi_0'(r)Z(r,\theta)
    \delta^\mu_r
    +
    \mathcal O(\varepsilon_q^4),
    \label{eq:app_Jext_vector_full}
\end{equation}
because
\begin{equation}
    \nabla_\mu J_{\rm ext}^{\mu}
    =
    \frac{2\tilde\beta\varepsilon_q^2}{r^2}
    \frac{d}{dr}
    \left[
        r^2N^2(r)\phi_0'(r)Z(r,\theta)
    \right]
    +
    \mathcal O(\varepsilon_q^4).
    \label{eq:app_Jext_divergence_full}
\end{equation}
For the static spacelike constant-$X$ branch, we have that
\begin{equation}
    N^2(r)\big(\phi_0'(r)\big)^2
    =
    -2X_\star
    \equiv C>0,
    \quad
    \phi_0'(r)=\sigma\frac{\sqrt C}{N(r)},
    \quad
    \sigma=\pm1.
    \label{eq:app_phi0prime_static}
\end{equation}
If $J_{\rm ext}^{\mu}=0$, the quadrupole equation gives the first integral
\begin{equation}
    r^2N^2(r)\phi_0'(r)\mathcal X_{2,2}(r)
    =
    Q_2,
    \label{eq:app_X22_first_integral}
\end{equation}
where $Q_2$ is an integration constant.  Using Eq.~\eqref{eq:app_phi0prime_static}, we find an expression for the quadrupolar part at second order as
\begin{equation}
    \mathcal X_{2,2}(r)
    =
    \frac{Q_2}{\sigma\sqrt C\,r^2N(r)}.
    \label{eq:app_X22_sourcefree_solution}
\end{equation}
This expression diverges at a nonextremal black-hole horizon unless $Q_2=0$.  Hence the regular sourceless radial limit forces
\begin{equation}
    \mathcal X_{2,2}=0,
    \qquad
    \Pi_2=0,
    \qquad
    S_K=0,
    \label{eq:app_sourcefree_X22_zero}
\end{equation}
and with homogeneous boundary conditions, Eq.~\eqref{eq:HK_system_app} then gives $H=K=0$.  The sourceless radial limit therefore does not generate the circular deformation which can lead to CTCs.

For the non-zero charge branch we choose a smooth localized profile as
\begin{equation}
    \mathcal X_{2,2}^{\rm src}(x)
    =
    \tilde C_X
    \frac{x_b^4}{x^4}
    \left(
        1-\frac{x_b}{x}
    \right)^2,
    \qquad
    \tilde C_X\neq0,
    \label{eq:X2_profile_nz}
\end{equation}
where $x_b$ is the black-hole root of $N^2$ and the dimensionless number $C_X$ controls the normalized source amplitude defined in Eq.~\eqref{eq:CX_simple_def}.  Since $m_{\rm src}$ is not identically zero, Eqs.~\eqref{eq:app_SK_def_nzflux} and \eqref{eq:app_Pi2_def_nzflux} imply nonzero $\Pi_2$ and $S_K$ on the support of this profile.

\subsection{AdS exterior boundary-value problem}

For $\lambda<0$, we solve the source-response system outside the positive root $x_b$ of $N^2(x)=0$.  Near $x_b$, define $y=x-x_b$.  The denominator in $c_{HH}$ then obeys
\begin{equation}
    6-3x+\lambda x^3
    =
    -3xN^2(x),
\end{equation}
so that, at a nondegenerate horizon,
\begin{equation}
    c_{HH}(x;\lambda)
    =
    -\frac{1}{y}
    +
    \frac{1}{2(3-x_b)}
    +
    \mathcal O(y).
    \label{eq:cHH_horizon_expansion_ads}
\end{equation}
The remaining coefficients are finite, and
\begin{equation}
    a_{KK,b}
    =
    \frac{2}{3-x_b}.
    \label{eq:aKK_horizon_ads}
\end{equation}
Writing
\begin{equation}
    H=H_b+H_1y+\mathcal O(y^2),
    \qquad
    K=K_b+K_1y+\mathcal O(y^2),
\end{equation}
the pole in the $H'$ equation is $-H_b/y$.  Regularity therefore gives
\begin{equation}
    H_b=0.
    \label{eq:H_horizon_regular_ads}
\end{equation}
The finite terms give
\begin{equation}
    H_1=-\frac12(a_{KK,b}K_b+S_b),
    \qquad
    K_1=a_{KK,b}K_b+S_b,
    \label{eq:horizon_derivs_general_ads}
\end{equation}
where $S_b=S_K(x_b;\lambda)$.  For the localized profile in Eq.~\eqref{eq:X2_profile_nz}, $S_b=0$, so the initial data at $x_0=x_b+\epsilon_h$ are
\begin{equation}
    H(x_0)
    =
    -\frac{K_b}{3-x_b}\epsilon_h,
    \qquad
    K(x_0)
    =
    K_b
    \left(
        1+\frac{2\epsilon_h}{3-x_b}
    \right).
    \label{eq:ads_frobenius_initial_data}
\end{equation}
The remaining shooting parameter $K_b$ is fixed by the outer condition
\begin{equation}
    H(x_{\rm out})=0,
    \qquad
    K(x_{\rm out})=0,
    \label{eq:ads_outer_conditions}
\end{equation}
in a least-squares sense for a sufficiently large $x_{\rm out}$.  In the analysis, we accept a point only when the outer residual is below a certain tolerance as stated in the main text.

\subsection{dS static-patch boundary-value problem}

For $0<\lambda<1/9$, the source-response system has a black-hole root $x_b$, a cosmological root $x_c$, and an internal point
\begin{equation}
    x_s
    =
    \left(\frac{3}{\lambda}\right)^{1/3},
    \qquad
    x_b<x_s<x_c,
    \label{eq:xs_ds_def}
\end{equation}
where $3-\lambda x_s^3=0$.  This internal point is a regular singular point of the radial source-response system.  Let $Q(x)=3-\lambda x^3$ and $y=x-x_s$.  Since $Q'(x_s)=-3\lambda x_s^2$, the pole residues are
\begin{align}
    a_{KK}
    &=
    \frac{B_s}{y}+\mathcal O(1),
    &
    B_s
    &=
    \frac{6}{Q'(x_s)}
    =
    -\frac{2x_s}{3},
    \label{eq:Bs_residue_ds}
    \\
    a_{KH}
    &=
    \frac{A_s}{y}+\mathcal O(1),
    &
    A_s
    &=
    \frac{2(6+6x_s+\lambda x_s^3)}{x_sQ'(x_s)}
    =
    -\left(2+\frac{4x_s}{3}\right),
    \label{eq:As_residue_ds}
    \\
    S_K
    &=
    \frac{S_s}{y}+\mathcal O(1),
    &
    S_s
    &=
    -\frac{\Pi_2(x_s)}{4\lambda}
    =
    \frac{\mathcal X_{2,2}(x_s)m_{\rm src}(x_s;\lambda)}{\lambda}.
    \label{eq:Ss_residue_ds}
\end{align}
The pole part of Eq.~\eqref{eq:HK_system_app} is therefore
\begin{align}
    H'
    &=
    -\frac{A_sH+B_sK+S_s}{y}
    +
    \mathcal O(1),
    \label{eq:H_pole_ds}
    \\
    K'
    &=
    \frac{A_sH+B_sK+S_s}{y}
    +
    \mathcal O(1).
    \label{eq:K_pole_ds}
\end{align}
A finite differentiable solution through $x_s$ must satisfy the algebraic compatibility condition
\begin{equation}
    A_sH(x_s)+B_sK(x_s)+S_s=0.
    \label{eq:internal_regular_condition_ds}
\end{equation}

At each nondegenerate static-patch horizon $x_i\in\{x_b,x_c\}$, the same Frobenius analysis as in the AdS case gives
\begin{equation}
    H_i=0,
    \qquad
    H_1=-\frac12(a_{KK,i}K_i+S_i),
    \qquad
    K_1=a_{KK,i}K_i+S_i,
    \qquad
    a_{KK,i}=\frac{2}{3-x_i}.
    \label{eq:ds_horizon_regular_data}
\end{equation}
For the localized profile, $S_b=0$ at the black-hole horizon.  At the cosmological horizon $S_c=S_K(x_c;\lambda)$ is generally nonzero.  We can therefore solve the dS finite-patch problem on the two intervals
\begin{equation}
    [x_b+\epsilon_h,x_s-\epsilon_s],
    \qquad
    [x_s+\epsilon_s,x_c-\epsilon_h].
\end{equation}
The unknown horizon amplitudes are $K_b$ and $K_c$.  The left and right integrations are matched by imposing
\begin{align}
    H_L(x_s)&=H_R(x_s),
    \label{eq:ds_match_H}
    \\
    K_L(x_s)&=K_R(x_s),
    \label{eq:ds_match_K}
    \\
    A_s\bar H_s+B_s\bar K_s+S_s&=0,
    \label{eq:ds_match_internal}
\end{align}
where $\bar H_s=(H_L+H_R)/2$ and $\bar K_s=(K_L+K_R)/2$.

The response $x_b$ and $x_c$ govern the behaviour of the radial ODE.  The radial domain used for the chronology diagnostic is taken from the exact zero-charge Kerr--(A)dS geometry. For a given $(\lambda,\chi)$, the rotating horizons are the positive roots of $\Delta_x=0$, with $\Delta_x$ defined in Eq.~\eqref{eq:dimless_kerrads_functions}.  For $\lambda\le0$, the exterior starts at the largest positive root $x_+$.  For $\lambda>0$, the static-patch domain is $x_+<x<x_{c,{\rm rot}}$, where $x_{c,{\rm rot}}$ is the largest positive root of the same equation.  Only points satisfying $\Xi>0$ and $\Delta_\theta>0$ are included.  We interpolate $H$ and $K$ onto this domain and insert them into Eqs.~\eqref{eq:charged_gtautau_exactspin}--\eqref{eq:charged_gphiphi_exactspin}.  The zero-charge domain is exact in $\chi$, while the charged circular response remains the leading $\mathcal O(\chi^2\varepsilon_q^2)$ Hartle--Thorne correction.

\subsection{Numerical implementation and perturbative control}
\label{app:numerical_implementation}

In this subsection, we outline the numerical implementation used for Fig.~\ref{fig:alpha_beta_ctc_scan}.  All quantities are dimensionless and use the variables in Eq.~\eqref{eq:x_chi_lambda_def}.  For each point in the signed $(\alpha,\tilde\beta)$ plane we first map to
\begin{equation}
    \lambda
    =
    \Lambda_{\rm eff}(GM)^2
    =
    \frac{\kappa\Mstar^4(GM)^2}{4}
    \frac{\alpha^2}{\tilde\beta} .
    \label{eq:lambda_alpha_beta_numeric}
\end{equation}
The static spacelike constant-$X$ branch requires $\alpha\tilde\beta>0$.  Points with $\alpha\tilde\beta<0$ are therefore marked as off branch.  The low-gradient cutoff check is implemented as
\begin{equation}
    \frac{|X_{c\star}|}{\Mstar^4}
    =
    \frac{1}{2|B|}
    =
    \frac{\alpha^2}{2|\tilde\beta|}
    <
    X_{\rm max},
    \label{eq:Xstar_numeric_cut}
\end{equation}
where $X_{\rm max}$ is the displayed cutoff threshold.  Points that fail Eq.~\eqref{eq:Xstar_numeric_cut} are not used to declare a controlled chronology violation.

The radial response problem and the Kerr--(A)dS diagnostic domain use different radial functions.  The response singular points are the roots of the static lapse $N^2(x)$ in Eq.~\eqref{eq:static_lapse_exactspin_source}; the chronology domain is instead determined by the roots of $\Delta_x$ defined in Eq.~\eqref{eq:dimless_kerrads_functions}.  We therefore compute the result in two stages, first solving Eq.~\eqref{eq:HK_system_app} on the appropriate interval, and then interpolating $H$ and $K$ onto the Kerr--(A)dS diagnostic domain determined by $\Delta_x$ and taking the result to leading order in $\mathcal O(\chi^2\varepsilon_q^2)$.

For $\lambda<0$ we solve the AdS-like branch by one-sided shooting from $x_b+\epsilon_h$ to a large cutoff $x_{\rm out}$.  The horizon data are the Frobenius data in Eq.~\eqref{eq:ads_frobenius_initial_data}.  The single shooting amplitude $K_b$ is chosen by least squares to minimize
\begin{equation}
    {\cal R}_{\rm AdS}
    =
    \left[
        H(x_{\rm out})^2+K(x_{\rm out})^2
    \right]^{1/2} .
    \label{eq:R_ads_numeric}
\end{equation}
The point is retained for diagnostic purposes only if the integration succeeds and ${\cal R}_{\rm AdS}$ is below the tolerance specified for the scan.  For $0<\lambda<1/9$ the dS-like branch is solved by the split static-patch boundary-value problem (BVP) described in Sec.~\ref{app:nzflux_technical}.  The numerical residual vector is
\begin{equation}
    {\bf R}_{\rm dS}
    =
    \begin{pmatrix}
        H_L(x_s)-H_R(x_s)\\
        K_L(x_s)-K_R(x_s)\\
        A_s\bar H_s+B_s\bar K_s+S_s
    \end{pmatrix},
    \label{eq:R_ds_numeric}
\end{equation}
with $\bar H_s=(H_L+H_R)/2$ and $\bar K_s=(K_L+K_R)/2$ in the finite-offset implementation.  The two shooting amplitudes $(K_b,K_c)$ are chosen by least squares, and the point is accepted only if $\|{\bf R}_{\rm dS}\|$ is below the stated tolerance.  We mark Failed BVP points as unsolved.

After interpolation onto the Kerr--(A)dS diagnostic domain $D(\lambda,\chi)$, we evaluate the minima ${\cal G}$ and ${\cal T}$ defined in Eq.~\eqref{eq:G_min_exactspin}.  The condition ${\cal G}<0$ signals a possible circular CTC.  The condition ${\cal T}<0$, or equivalently $\Delta_{\rm K}>0$ somewhere in the domain, signals a possible breakdown of the Killing-sector time function.

A zero crossing of the truncated expression for $g_{\varphi\varphi}$ is declared controlled only if the charged perturbation remains small. For the minimal ansatz in Eqs.~\eqref{eq:hq_tautau_exactspin}--\eqref{eq:hq_phiphi_exactspin}, the control variable is
\begin{equation}
    {\cal C}_{\rm amp}
    =
    \max_{x\in D}
    \left\{
        \left|
        \varepsilon_q^2\chi^2P_2(\cos\theta)H(x)
        \right|,
        \left|
        \varepsilon_q^2\chi^2P_2(\cos\theta)K(x)
        \right|
    \right\},
    \label{eq:C_amp_def}
\end{equation}
where the maximum excludes the small horizon buffers used in the integration.  The direct azimuthal-component control variable is
\begin{equation}
    {\cal C}_{\varphi\varphi}
    =
    \max_{x\in D}
    \left|
    \frac{
        \varepsilon_q^2x^2\sin^2\theta\,\chi^2K(x)P_2(\cos\theta)
    }{
        \bar g_{\varphi\varphi}(x,\theta)
    }
    \right| .
    \label{eq:C_gphiphi_def}
\end{equation}
The strict perturbative evaluation uses
\begin{equation}
    {\cal C}_{\rm amp}<c_{\rm max},
    \qquad
    {\cal C}_{\varphi\varphi}<c_{\rm max},
    \qquad
    c_{\rm max}=0.3 .
    \label{eq:control_cut_numeric}
\end{equation}
This value is permissive for a perturbative expansion; a more conservative production scan can replace it by $c_{\rm max}=0.1$.  Points with ${\cal G}<0$ that fail Eq.~\eqref{eq:control_cut_numeric} are labeled as possible circular CTCs outside perturbative control.  

Setting $\varepsilon_q=0$ reproduces the exact Kerr--(A)dS identities \eqref{eq:kerrads_det_identity} and \eqref{eq:kerrads_gphiphi_positive_identity_dimless}, and hence gives no exterior circular CTCs in the usual domain.  Setting $\mathcal X_{2,2}=0$ gives $S_K=0$ and returns the homogeneous solution $H=K=0$ with the boundary conditions used above.  Decreasing the dimensionless source amplitude $C_X$ decreases $H$ and $K$ linearly and removes the possible zero crossings before any controlled circular CTC appears.

\section{Residual stress for the sourced branch}
\label{app:exact_source_closure}

The derivative current in Eq.~\eqref{eq:source_action} fixes the scalar equation but not the stress tensor of the external sector.  The residual external stress energy can be fixed by solving the inverse problem by writing the Einstein equations as
\begin{equation}
    T^{\rm ext}_{\mu\nu}[g,\phi]
    =
    \frac{1}{\kappa}G_{\mu\nu}[g]
    -P_X\nabla_\mu\phi\nabla_\nu\phi
    -g_{\mu\nu}P(X).
    \label{eq:app_residual_Text_exact}
\end{equation}

The required force balance follows from the scalar equation.  Writing the covariant derivative of the stress-energy tensor as
\begin{equation}
    \nabla^\mu T^{(\phi)}_{\mu\nu}
    =
    \left[\nabla_\mu(P_X\nabla^\mu\phi)\right]\nabla_\nu\phi
    =
    \mathcal J_{\rm ext}\nabla_\nu\phi,
    \label{eq:app_scalar_stress_div_final}
\end{equation}
where \(\mathcal J_{\rm ext}=\nabla_\mu J_{\rm ext}^\mu\), we then find from the Bianchi identity then gives
\begin{equation}
    \nabla^\mu T^{\rm ext}_{\mu\nu}
    =
    -\mathcal J_{\rm ext}\nabla_\nu\phi .
    \label{eq:app_Text_balance_exact}
\end{equation}
At \(\mathcal O(\varepsilon_q^2)\), after subtracting the zero-charge vacuum energy and projecting onto the Hartle--Thorne sector, the linearized Einstein equation is
\begin{equation}
    \mathcal E_{\mu\nu}^{\rm HT,\lambda}[h^{(q)}]
    =
    \kappa\left(\tau^{(\phi)}_{\mu\nu}+\tau^{\rm ext}_{\mu\nu}\right),
    \qquad
    \mathcal E_{\mu\nu}^{\rm HT,\lambda}[h]
    \equiv
    \delta G_{\mu\nu}[h]+\Lambda_{\rm eff}h_{\mu\nu},
    \label{eq:app_linearized_complete_Einstein}
\end{equation}
with
\begin{equation}
    \tau^{(\phi)}_{\mu\nu}
    =
    2\tilde\beta\left[\mathcal X_{2,0}(x)+\mathcal X_{2,2}(x)P_2(\cos\theta)\right]
    \nabla_\mu\phi_\star\nabla_\nu\phi_\star .
    \label{eq:app_tau_phi_q2}
\end{equation}
Thus the residual perturbative source required is
\begin{equation}
    \tau^{\rm ext}_{\mu\nu}
    =
    \frac{1}{\kappa}\mathcal E_{\mu\nu}^{\rm HT,\lambda}[h^{(q)}]
    -\tau^{(\phi)}_{\mu\nu}.
    \label{eq:app_tau_ext_linear_residual}
\end{equation}
\bibliography{sample}

\end{document}